\documentclass[doublecolumn]{IEEEtran}
\usepackage{cite}
\usepackage{amsmath,amssymb,amsfonts,bm}
\usepackage{graphicx}
\usepackage{textcomp}
\usepackage[table]{xcolor}
\usepackage{tcolorbox}   
\usepackage{color}
\usepackage{framed}
\usepackage{subcaption}
\usepackage{diagbox}
\usepackage[noend]{algpseudocode}
\usepackage{algorithmicx,algorithm}
\usepackage{tabularx}
\usepackage{colortbl}
\usepackage{listings}
\usepackage{booktabs}
\usepackage[noend]{algpseudocode}

\usepackage[colorlinks=true, citecolor=black, linkcolor=black]{hyperref}

\newcommand{\bc}[1]{\mbox{\boldmath $\mathcal{#1}$}}
\newcommand{\bs}[1]{\boldsymbol{#1}}
\newcommand{\mf}[1]{\mathbf{#1}}
\newcommand{\mb}[1]{\mathbb{#1}}
\newcommand{\F}{\mathrm{F}}

\newcommand{\T}{\mathrm{T}}

\begin{document}
\title{FieldFormer: Self-supervised Reconstruction of Physical Fields via Tensor Attention Prior}

\author{Panqi Chen, Siyuan Li, Lei Cheng, {\it Member, IEEE}, Xiao Fu, {\it Senior Member, IEEE}, Yik-Chung Wu,  {\it Senior Member, IEEE}, and Sergios Theodoridis, {\it Life Fellow, IEEE}
	
	\thanks{The work of Lei Cheng was supported in part by the National Natural Science Foundation of China under Grant 62371418. The work of Xiao Fu was supported in part by   the National Science Foundation (NSF) under Projects NSF ECCS-2024058 and NSF CCF-2210004. (Corresponding author: Lei Cheng)}
	\thanks{Panqi Chen, Siyuan Li, Lei Cheng are with the College of Information Science and Electronic Engineering, Zhejiang University, Hangzhou, China. 
		(e-mails: panq$\_$chen, 3180100878, lei$\_$cheng$\}$@zju.edu.cn); Xiao Fu is with the School of Electrical Engineering and Computer Science, Oregon State University, Corvallis, OR 97331 USA (e-mail: xiao.fu@oregonstate.edu); Yik-Chung Wu is with the Department of Electrical
		and Electronic Engineering, The University of Hong Kong, Hong Kong (email: ycwu@eee.hku.hk); Sergios Theodoridis is with the
 Department of Informatics and Telecommunications, 
 National and Kapodistrian University of
		Athens, Greece (email: stheodor@di.uoa.gr).}
	\thanks{This paper has supplementary downloadable material available at http://ieeexplore.ieee.org., provided by the author.}
}

\markboth{\quad}{\quad}%

\IEEEpubid{}

\maketitle

\begin{abstract}
Reconstructing physical field tensors  from  \textit{in situ} observations, such as radio maps and ocean sound speed fields, is crucial for enabling  environment-aware decision making in various applications, e.g., wireless communications  and underwater acoustics.  Field data reconstruction is often challenging, due to the limited and noisy nature of the observations, necessitating the incorporation of prior information to aid the reconstruction process. 
Deep neural network-based data-driven structural constraints (e.g., ``deeply learned priors'') have showed promising performance.  However, this family of techniques faces challenges such as model mismatches between training and testing phases. This work introduces FieldFormer, 
a self-supervised neural prior   learned solely from the limited \textit{in situ} observations without the need of offline training. Specifically,  the proposed framework starts with modeling the fields of interest using the tensor Tucker model  of a high multilinear rank, which ensures a universal approximation property for all fields. In the sequel, 
an attention
mechanism   is incorporated to learn the sparsity pattern that underlies the core tensor in order to reduce   the solution space.
 In this way, a ``complexity-adaptive'' neural representation,  grounded in the Tucker decomposition,  is obtained that can flexibly represent
 various types of fields.
A theoretical analysis is provided to   support the recoverability of the proposed  design. Moreover, extensive experiments, using various physical field tensors, demonstrate the superiority of the proposed approach compared to state-of-the-art baselines. The code is available at 
\it{\href{https://github.com/OceanSTARLab/FieldFormer}{ https://github.com/OceanSTARLab/FieldFormer}}.
	
\end{abstract}


\section{Introduction}

\IEEEPARstart{T}{he} 
 accurate characterization of  signal propagation in complex environments, such as underwater acoustics in the sea or electromagnetic waves in urban areas, is the stepping stone towards {\it environment-aware} wireless communications, target detection and recognition\cite{Radio,radio_spm, spline, hooi, TVT, TAES}. To accomplish this, several types of three-dimensional (3D) physical fields have been developed to provide valuable information across a given geographical region. Examples include the ocean sound speed field\cite{sound}, which governs sound transmission in a spatially 3D ocean environment, and the radio map\cite{radio_spm}, which reveals information about the propagation of radio power across two spatial domains and one frequency domain.

Despite the vital role of the aforementioned 3D physical fields, crafting a finely detailed field that precisely captures the rapid variations of physical quantities (such as sound speeds or radio powers) across multiple domains (such as spaces or frequencies) presents a highly challenging task. Due to the high cost of in-situ measurements, sensors are often sparsely deployed across the geographical region, leaving a substantial portion of the physical fields unobserved\cite{luo, natureoat}. Using such limited and potentially noisy samples to reconstruct the complete 3D physical field is a typical ill-posed inverse problem, which has undergone extensive studies in recent years\cite{Radio,radio_spm, spline, hooi,TNN,  BTD,  kri, kri2, DL_rm, AE_rm, sensor}.

Within the vast literature, the primary idea is to supplement the ill-posed reconstruction process with various informative priors of the associated physical fields. Early studies utilized {\it hand-crafted priors} rooted in basic assumptions about these fields, such as local smoothness\cite{spline, spline1,TV,kri,kri2} and global coherence\cite{LRTC,BGCM}. These assumptions could be readily translated into analytical forms such as total variations\cite{TV} and low-rank modeling\cite{BGCM,LRTC}. Despite their simplicity and interpretability, methods based on  hand-crafted priors  encounter challenges when the underlying structure of physical fields becomes complex. For instance, in the deep ocean, the presence of internal waves and eddies causes significant fluctuations in sound speed across large spatial scales\cite{sound}. Similarly, in urban areas, the proliferation of obstacles exacerbates the shadowing effect\cite{Radiomap}.


To excel in complex environments,  there has been a notable focus on {\it data-driven priors.} These priors can be broadly classified into two categories: supervised priors and unsupervised priors (also called trained priors and untrained priors, respectively). Supervised priors rely on training data from historical measurements or simulators.  
Despite their promising performance in applications like ocean sound speed field recovery \cite{K-svd, TDL} and radio map estimation \cite{ICC, Radio, AE_rm, sensor, quant}, these methods face a number of challenges:  1) the performance deteriorates substantially when the fields targeted for reconstruction follow different data distributions compared to the training data; 2) the learned priors need to be retrained if the scenarios change; and 3)  obtaining high-quality training data is not always feasible, especially for physical fields situated in deep-sea or rural areas \cite{luo, natureoat}.
To address the aforementioned challenges, there has been a surging interest in unsupervised data-driven priors \cite{unn}. These neural network-based priors only leverage inherent inductive-biases\cite{inductivebias} that match the structure  of the  data,  requiring no additional training data. This idea was used in various fields, e.g., image restoration\cite{DIP} and sound speed field recovery\cite{TNN}. Nonetheless, their effectiveness is often hampered by the limited number of observations and the predetermined model architecture/complexity. This prompts an intriguing question: {\it Can a self-supervised learning approach be devised to further distill knowledge from a limited amount of samples, which will lead to  a ``complexity-adaptive" data-driven prior to enhance physical field reconstruction? }

\begin{figure}[t]
	\centering
\includegraphics[width=1.01\linewidth]{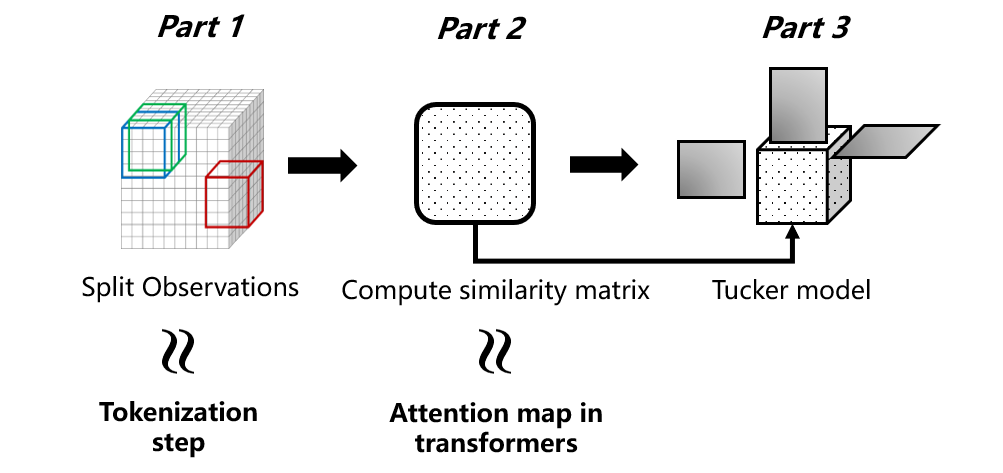}
	\caption{The schematic figure illustrating the  rationale of the proposed method.}
	\label{fig:sp}
\end{figure}

\noindent
{\bf Contributions.}
To address this question, we propose {\it FieldFormers}. These comprise neural  representations for field data that build upon  a) a tensor Tucker model, b) an attention mechanism and c) self-supervised learning. Central to our idea is to leverage the notion of attention mechanism
 to automatically adjust the complexity of the adopted Tucker model, so that the representation strikes a reasonable balance between universality and parsimony.  The main rationale of the proposed method lies in the following concept. We split the observation tensor in smaller cubes (see the first part of Fig.~\ref{fig:sp}). One could view this process as the equivalent of the tokenization step in natural language processing (NLP).   Then, a similarity matrix among the various tokens/cubes is computed, similar to the attention map in transformers (see the second part of Fig.~\ref{fig:sp}). The similarity (attention) matrix implicitly implies a sparsity structure, since less similar parts/tokens lead to low-value attention weights. It is exactly this information that will be exploited by imposing it on the adopted  Tucker model (see the third part of Fig.~\ref{fig:sp}).
Following this rationale, we devise (multi-head) tensor attention priors ((MH)TAP)  to enable learning the sparse patterns of the core tensor of an over-complete Tucker model, which will be elaborated in Sec.~\ref{sec:3}. The tensor attention mechanism is critical in capturing both short- and long-range dependencies among different areas of the field, and mapping such dependencies into the core tensor.
 
In addition to model design, the paper also studies  various aspects that are of theoretical interest.
We analyze the expected number of the non-zero elements in the core tensor to represent field data. Furthermore,  we provide recoverability guarantees under the proposed model, which reveal the trade-off between sample and model complexities.
Extensive experimental results, using ocean sound speed fields and radio maps, are presented that demonstrate the excellent performance of the proposed approaches.

\noindent
{\bf Notations:} Lower- and upper-case bold letters (e.g., $\mf x$ and 
$\mf X$) are used to denote vectors and matrices, respectively.  Upper-case bold calligraphic letters   and  upper-case calligraphic letters  (e.g., $\bc{X}$ and $\mathcal{X}$) are used to denote tensors and sets.  Operations $\otimes, \odot, \ast, \circ$ denote 
Kronecker product, Khatri-Rao product, Hadamard  product and outer product respectively. $\|\cdot\|_{\F}, \|\cdot\|_{0}$, $\|\cdot\|_{2}$ and $\|\cdot\|_{\ast}$ represent Frobenius norm, $L_0$ norm, $L_2$ norm and nuclear norm, respectively.  $|\mathcal{X}|$ represents the cardinality of set $\mathcal{X}$. $\mathbin{\|}$ and $\mod$ are exact division and modulus operators.



\section{Problem Statement and Prior Art}
\label{sec:2}

In this section, we present the problem setup of  3D physical fields reconstruction and introduce the prior art.

\subsection{Problem Setup}
The objective is to reconstruct the ground-truth 3D physical field, denoted as $\bc X_{\natural} \in \mathbb{R}^{I_1 \times I_2 \times I_3}$, from a limited number of noisy observations, denoted as $\bc Y \in \mathbb{R}^{I_1 \times I_2 \times I_3}$. The typical observation or sensing model is defined as follows:
\begin{equation}
	\label{om}
	\bc{Y} = \bc{O}\ast (\bc{X}_{\natural} + \bc{N}),
\end{equation}
where $\bc{N} \in \mathbb{R}^{I_1 \times I_2 \times I_3}$ represents the noise tensor, and the binary tensor $\bc{O}$ indicates the observed entries, where $\bc{O}(i_1,i_2,i_3)=1$ if the $(i_1,i_2,i_3)$-$th$ point is observed, and $\bc{O}(i_1,i_2,i_3)=0$ otherwise.

Based on the sensing model in~\eqref{om}, the reconstruction problem can be formulated in the following conceptual form: 
\begin{equation}
	\label{general_loss}
	\begin{split}
		&\min_{\bc{X}} ~\|\bc{Y} - \bc{O}\ast \bc{X} \|_{\F}^{2},\\
		&\text{s.t.}~~\bc{X} \in  \mathcal{F},
	\end{split}
\end{equation}
where $\mathcal{F}$ denotes a set of structural constraints of the 3D field $\bc{X}$. 
The recovery problem is ill-posed, as the number of observations is often much smaller than the signal dimension $I_1I_2I_3$. 
The key to tackling such a challenging  inverse task lies in selecting a proper  $\mathcal{F}$ that reflects prior information of ${\bc X}$ and incorporating the related information to recover ${\bc X}$. We also denote full observation $\tilde{\bc{Y}}$ as: 
\begin{equation}
	\label{om}
	\tilde{\bc{Y}} =  \bc{X}_{\natural} + \bc{N}.
\end{equation}

In the following subsections, we briefly review the prior art on designing $\mathcal{F}$ and the remaining challenges.

\subsection{Prior Art and Challenges Ahead}
\noindent
\textbf{Handcrafted prior:} Many early methods in this domain use a relatively simple constraint set ${\cal F}$, e.g., low (matrix/tensor) rank \cite{LRTC, BGCM}, and total variation \cite{TV,spline, spline1}. The respective implementation is also relatively straightforward: one can often approximate these constraints using convex regularization terms, e.g., using the tensor nuclear norm \cite{Tennorm}  $\frac{1}{3}\sum_{l=1}^{3}\|\mf{X}_{(l)}\|_{\ast}$ to approximate the low-rank constraint on ${\bc X}$, where $\mf{X}_{(l)}$ is the mode-$l$ folding of $\bc{X}, \forall l$. Although these constraints/regularization terms are simple to incorporate and easy to interpret, they often have limited capabilities in handling complex scenarios, requiring careful design for specific tasks. 



\noindent
\textbf{Supervised and unsupervised data-driven priors:} Another idea is to learn a generative model of ${\bc X}$ from historical or simulated data $\{\bc{X}^{\rm train}_n\}_{n=1}^N $. This can be done via using popularized neural generative models such as autoencoders (AEs) \cite{Radio} or generative adversarial networks (GANs) \cite{quant}; e.g., conceptually, AE-based generative model learning can be formulated as
\begin{align}
    \min_{\bm \theta}~\frac{1}{N}\sum_{n=1}^N\| D_{\bm  \theta}(Q_{\bm  \beta}(\bc{X}^{\rm train}_n)) - \bc{X}^{\rm train}_n) \|_{\rm F}^2.
\end{align}
Here, the representation model ${\bc X}\approx D_{\bm \theta}(\bm z)$, where
$ D_{\bm \theta}(\cdot)$ denotes a neural generative model and $\bm z$ is the associated latent representation of ${\bc X}$; $Q_{\bm  \beta}({\bc X})$ is the so-called ``encoder'' parameterized by $\bm  \beta$ such that $Q_{\bm  \beta}({\bc X})\approx {\bm z}$.
Once $D_{\bm \theta^{\star}}(\cdot)$ is learned with $\bs \theta^{\star}$ denoting the learned parameters, problem~\eqref{general_loss} can be simplified as
\begin{align}
    \min_{\bm z}~\|\bc{Y} - \bc{O}\ast  D_{\bm \theta^{\star}}(\bm z) \|_{\F}^{2}.
\end{align}
This type of data-driven prior partially mitigates the challenges related to handcrafted priors, particularly excelling in capturing the intricate details of physical fields\cite{ICC,Radio}.  
Nonetheless, their performance heavily depends on the quality and quantity of the training datasets $\{\bc{X}^{\rm train}_n\}_{n=1}^N$. Consequently, 
they face difficulties in adapting to real-time environmental shifts, potentially requiring re-training.

A workaround is to employ the so-called unsupervised priors, i.e., representing $\bc{X}$ as $\bc{X}=D(\bm \theta)$, i.e., a neural network with untrained parameters $\bm \theta$, and solve the following:
\begin{align}
    \min_{\bm \theta}~\|\bc{Y} - \bc{O}\ast  D(\bm \theta) \|_{\F}^{2}.
\end{align}
In this way, neural architectures introduce useful inductive bias to model complex $\bc{X}$, but training data is not needed. The input of the corresponding neural network is excited by random noise \cite{DIP, dp}.
Such untrained models attracted much attention from the vision community \cite{DIP, dp} and were also used in sound speed field recovery \cite{TNN}.  However, designing the neural architecture is often nontrivial, and implementing such complex models often requires many heuristics that are hard to interpret, e.g., early stopping \cite{DIP}.

\noindent
\textbf{Self-supervised approach:}
To address the limitations of both supervised and unsupervised methods mentioned earlier, this paper follows a different path aiming to reconstruct  physical fields using a self-supervised approach. The essence of  self-supervised learning  is   to learn the underlying structure of the data by generating features at the output of an encoder, which  is trained via targets that are constructed from the available  unlabeled  data, e.g., \cite{ml3,bengio2021deep, SSL}.   In our specific context, the reconstruction problem can be formulated as follows:	
\begin{equation}
	\label{loss_1}
	\begin{split}
		&\min_{\bs \theta} \|\bc{Y} - \bc{O}\ast \bc{X} \|_{\F}^{2},\\
		&\text{s.t.} \quad \bc{X} = D_{\bs \theta}(\bc{Y}).
	\end{split}
\end{equation} 
Here, $D_{\bs \theta}(\bc{Y})$  defines the feasible set $\mathcal{F}$ (the feasible set could be considered as the encoder branch in a self-supervised task, starting from the input $\bc Y$) and recovers the entire field from limited and noisy observations $\bc{Y}$, which act as the respective targets for the training.  The parameters $\bs \theta$ are optimized through self-supervised learning,  contrasting with supervised approaches with pre-trained parameters.  On the other hand, the proposed approach leverages input observations $\bc Y$ to dynamically learn the model architecture/complexity, while earlier unsupervised methods are reliant on a fixed neural network architecture/complexity. Self-supervised prior learning was recently seen in vision \cite{DAP} and hyperspectral imaging \cite{luo2022self}, showing appealing characteristics.
 {However, \cite{DAP} and \cite{luo2022self} are not well suited for handling complex physical fields, as the former fails to capture multidimensional interactions within 3D fields, while the latter imposes restrictive low-rank constraints.}
In this work, our interest lies in designing self-supervised prior learning mechanisms tailored for 3D physical fields.

\section{Proposed Approach}
\label{sec:3}

In this section, we first introduce the tensor Tucker model and the attention mechanism as preliminaries, and then we elaborate how we propose our framework based on them.


\subsection{Preliminaries}
\label{s1}
\begin{figure}[!t]
	\centering
	\includegraphics[width=0.9\linewidth]{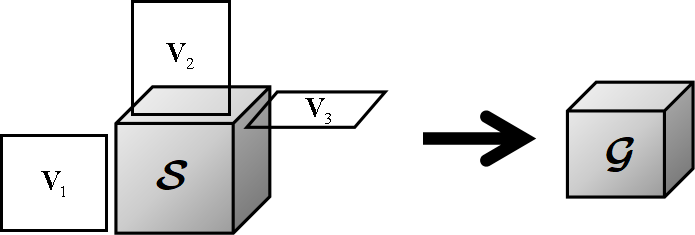}
	\caption{Illustration of a 3-$rd$ order tensor Tucker model. }
	\label{fig:TL}
\end{figure}

\noindent{\bf Tensor Tucker model:} The tensor Tucker model serves as the cornerstone of the proposed framework. {   We choose Tucker model because it appears to fit to the context of field estimation naturally. In particular,  the core tensor resembles the attention map that admits interesting interpretation. Other tensor decomposition models may not offer such an immediate connection to attention.} Specifically, we employ a third-order Tucker model in this work, as illustrated in Fig.~\ref{fig:TL}. Its mathematical expression is as follows:	
\begin{equation}
	\label{eq:tl}
	\bc{G} = \bc{S} \times_1 \mf{V}_1  \times_2 \mf{V}_2  \times_{3} \mf{V}_3,
\end{equation}	
where $\bc{S} \in \mb{R}^{R_1\times R_2 \times R_3}$  represents the core tensor and   $\{\mf{V}_l \in \mb{R}^{I_l\times R_l} \}_{l=1}^3$  denote three sets of factor matrices. The output  is denoted by $\bc{G}  \in \mb{R}^{I_1\times I_2 \times I_3}$.   
Note that $\times_l$ stands for mode-$l$ product. Specifically, the  mode-$l$ product of a third order tensor $\bc{A}\in \mb{R}^{I_1 \times \cdots \times I_l \times \cdots \times I_L }$ and a matrix $\mf{B} \in \mb{R}^{J_l \times I_l}$, denoted as $\bc{A} \times_l \mf{B}$, produces a $L$-$th$ order tensor $\bc{C}\in \mb{R}^{I_1 \times \cdots \times J_l \times \cdots \times I_L }$. And it can be  expressed as 
\begin{equation}
    \bc{C}_{i_1, i_2,\cdots, j_l, \cdots i_L } = \sum_{k=1}^{I_l} \bc{A}_{i_1, i_2,\cdots, k, \cdots i_L } \mf{B}_{j_l, k}.
\end{equation}

The Tucker model was introduced as \textit{higher-order singular value decomposition} (SVD) \cite{lieven} as it can retain orthogonality of $\bm V_l, \forall l$  (yet other decomposition such as the canonical polyadic decomposition (CPD) \cite{Tucker-als} cannot).
The Tucker model is a {\it universal representer}---that is, when $R_1,R_2$ and $R_3$ are large enough (up to $R_l=I_l$), any tensor can be expressed by a Tucker decomposition model\cite{Tucker-als}. This is analogous the matrix case---i.e., any real-valued matrix admits an SVD.


\noindent{\bf Attention mechanism:}  The scaled dot-product attention mechanism\cite{attention} has been widely adopted in NLP and computer vision\cite{bert, maecv} for its powerful feature extraction capabilities. 
Given $N$ input vectors of dimension $K$ (e.g., word embeddings), they can be organized into the input matrix $\mf{P}\in \mathbb{R}^{N \times K}$. The matrix $\mf{P}$ is then projected into different  ``embedding spaces'':
\begin{equation}
     \mf Q = \mf P \mf W_Q, \mf K = \mf P \mf W_K, \mf V = \mf P \mf W_V,
 \end{equation}
where $\mf W_Q, \mf W_K, \mf W_V \in \mb{R}^{K \times M}$ are the feature embedding matrices and $M$ represents the latent   embedding dimension.  The matrices $\mf{Q}, \mf{K},$ and $\mf{V}$ are the query, key, and value matrices, respectively,  all sharing the size of $N \times M$.
Consequently, the formulation of attention is as follows:
\begin{align}
	\label{eq:attention}
	\text{Attention}(\mf{Q}, \mf{K}, \mf{V}) = \text{SoftMax}(\frac{\mf Q \mf K^\T}{\sqrt{M}}) \mf V,
\end{align}
where SoftMax function does row normalization of  $\mf Q\mf K^{\T}$  after scaled by  $\sqrt{M}$. Note that $\mf Q \mf K^\T$ results in an attention map (i.e., row-by-row correlation matrix) of size $N \times N$, with the $n$-$th$ row representing the similarities between $n$-$th$ input  vector and the rest.
In a nutshell, attention  is basically an expansion of the input matrix $\mf{P}$ in terms of value vectors (i.e., rows in $\mf{V}$) with weighting coefficients expressing mutual similarities. Less similar vectors are weighted with small weights. Thus this mechanism can also be used to impose sparsity in the model.

\begin{figure}[t]
	\centering
	\begin{subfigure}{0.208\textwidth}
\includegraphics[width=\textwidth]{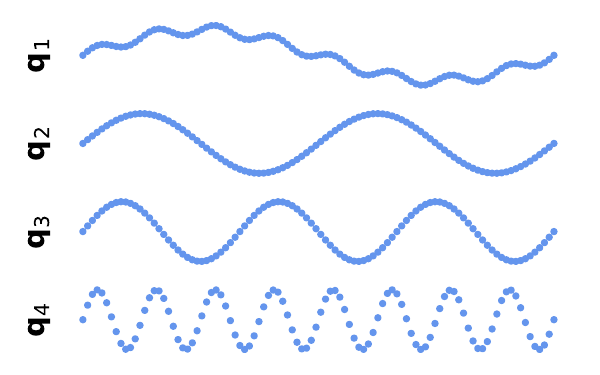}
		\caption{Toy example of query matrix $\mf{Q}$. The first row of $\mf{Q}$ is a sine wave in combination with $\mf{k}_1$ and $\mf{k}_4$ (as shown in (b)). The rest rows of $\mf{Q}$ are similar to that of $\mf{K}$.    }
		\label{fig:qsub1}
	\end{subfigure}
	\hspace{0cm}
	\begin{subfigure}{0.208\textwidth}
		\includegraphics[width=\textwidth]{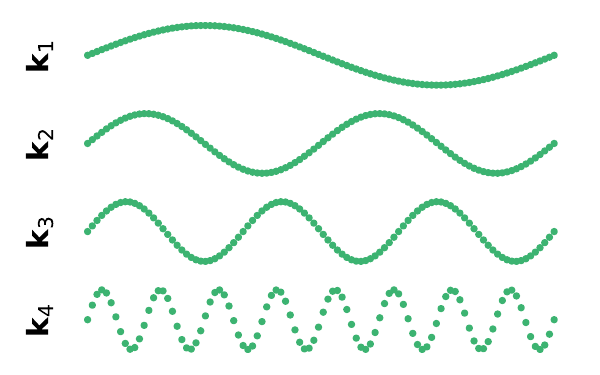}
		\caption{Toy example of key matrix $\mf{K}$.      Each row of $\mf{K}$ (i.e., $\mf{k}_l,  l=1,2,3,4$) is a sine wave with different frequencies. Therefore, they are orthogonal to each other.}
		\label{fig:ksub2}
	\end{subfigure}
 \begin{subfigure}{0.33\textwidth}
		\includegraphics[width=\textwidth]{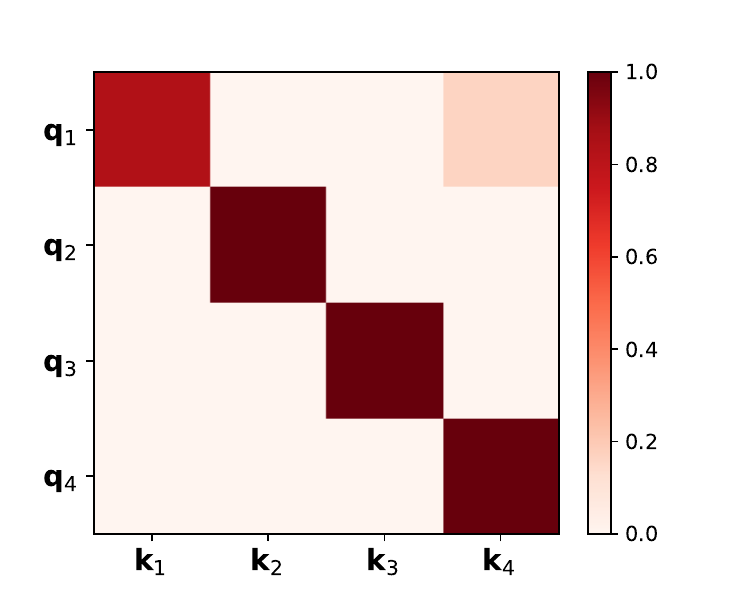}
		\caption{Toy example of $\mf{Q}\mf{K}^{\T}$. }
		\label{fig:sub3}
	\end{subfigure}
	\caption{Toy example of query, key matrices and their similarities represented by $\mf{Q}\mf{K}^{\T}$. Obviously, only $\mf{q}_1$  shows some similarities with $\mf{k}_1$ and $\mf{k}_4$ while others show no similarities but themselves.}
	\label{fig:qkt}
\end{figure}

A toy example is provided in Fig.~\ref{fig:qkt} to illustrate the process of computing an attention map, \textit{which  typically reflects some sparse patterns due to the predominantly incoherent nature of the latent features.}

\subsection{Proposed Sparse Tensor Attention Module}
\label{s2}
\begin{figure*}[!t]
	\centering
	\begin{minipage}{\textwidth}
		\centering
		\includegraphics[width=1.05\linewidth]{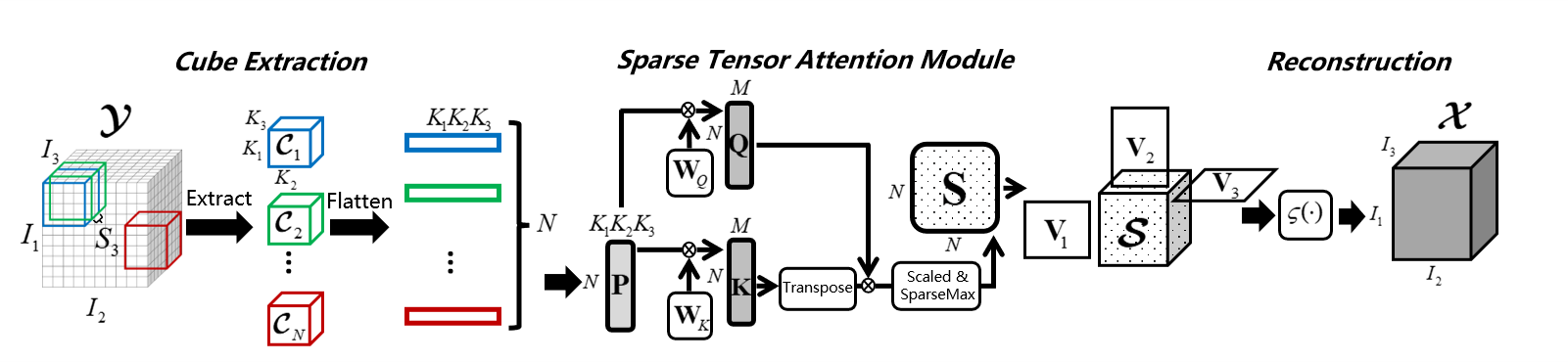} 
		\caption{ The detailed architecture of the proposed tensor attention prior (TAP) model for reconstructing 3D physical fields with limited observations.}
		\label{fig:sa}
		\vspace{-2mm}
	\end{minipage}
\end{figure*}
The expressiveness of the Tucker model depends on the sizes or dimensions of the core tensor $\bc{S} \in \mb{R}^{R_1\times R_2 \times R_3}$ in comparison to the output tensor $\bc{G} \in \mb{R}^{I_1\times I_2 \times I_3}$. Specifically, when the core tensor is  large, i.e., $R_l \ge I_l, \forall l$, it becomes expressive enough to represent an arbitrary tensor of size $(I_1, I_2, I_3)$. 
However, in the absence of suitable regularization, directly fitting such a Tucker  model to limited observations of physical fields can lead to overfitting. To address this issue, previous methods have proposed sparsifying the core tensor $\bc{S}$ by incorporating various sparsity-aware regularizers\cite{TDL, tong2023bayesian, TDL2}. Nevertheless, these handcrafted regularizers   do not adapt to {\it in-situ} data,  and thus often exhibit model mismatches in field estimation problems.

Towards data-adaptive sparse coding for the core tensor $\bc{S}$, we propose leveraging the attention mechanism. Our idea is inspired by the underlying similarities  between the expansions in Eq.~\eqref{eq:tl} and Eq.~\eqref{eq:attention}. 
By comparing the two expressions, one can interpret the Tucker model as a   tensorized version of the attention matrix. Specifically, the core tensor $\bc{S}$ acts as the weight coefficient tensor, and the three factor matrices $\{\mf{V}_l \}_{l=1}^3$ serve as the value matrices. 
This observation motivates us to propose a way for automatically learning the sparsity pattern of the core tensor $\bc S$.

\subsubsection{Local Region Representation} 
To achieve this goal, as shown in the first part of Fig.~\ref{fig:sa},  we begin by extracting cubes from the observed 3D physical field  $\bc{Y} \in \mb{R}^{I_1 \times I_2 \times I_3}$ using 3D windows of sizes $(K_1, K_2, K_3)$ and strides of sizes $(S_1, S_2, S_3)$ in the three modes\footnote{{ Guidelines for the selection of window sizes and stride sizes are presented in Appendix~L.}
}.  This process generates a total of $N$ cubes, where $N= \prod_{l=1}^{3} J_l$ and  $J_l = (\frac{I_l-K_l}{S_l}+1)$. Specifically, we define $\bc{C}_n\in \mathbb{R}^{K_1\times K_2 \times K_3}$ to represent the $n$-$th$ cube extracted from $\bc{Y}$:
\begin{equation}
\begin{split}
    &\bc{C}_n= \bc{Y}(\mf{i}_1,\mf{i}_2,\mf{i}_3),
\end{split}
\end{equation}
where the index set $\{\mf{i}_l = 1+(m_l-1)S_l:(m_l-1)S_l+K_l, \forall l\}$ 
and $m_1= (n-1) \| J_2J_3 + 1, m_2 = [(n-1) \mod J_2J_3] \| J_3 + 1, m_3 = (n-1) \mod J_3 +1$. 
A clearer illustration of patch extraction can be found in the cube extraction part of Fig.~\ref{fig:sa}, where the first cube $\bc{C}_1$ is represented in blue, the second cube $\bc{C}_2$ is in green, and so forth.
These $N$ cubes are then vectorized, resulting in a data matrix $\mathbf{P} =[{\bf c}_1,\ldots,{\bf c}_N]^{\T} \in \mathbb{R}^{N \times K_1 K_2 K_3}$ where ${\bf c}_n ={\rm vec}(\bc{C}_n) \in \mb{R}^{K_1 K_2 K_3}$.

\subsubsection{Sparse Tensor Attention Construction}
Next, we introduce the proposed sparse tensor attention  (STA) module. As shown in the second part of Fig.~\ref{fig:sa}, we project the data matrix $\mathbf{P}$ into two latent spaces using the embedding matrices $\mathbf{W}_Q, \mathbf{W}_K \in \mathbb{R}^{K_1K_2K_3 \times M}$. This projection yields the query matrix $\mathbf{Q} = \mathbf{P} \mathbf{W}_Q \in \mathbb{R}^{N \times M}$ and the key matrix $\mathbf{K} = \mathbf{P} \mathbf{W}_K \in \mathbb{R}^{N \times M}$. The dot products between the query and key matrices are then computed, resulting in a $N \times N$ correlation matrix $\mf{Q}\mf{K}^{\T}$. 

Unlike the traditional scaled dot-product  attention mechanism in \eqref{eq:attention}, which assumes the full observation and scales the correlation matrix  with a constant $\sqrt{M}$, we apply element-wise division with a scaling  matrix  $\mf{M}$. The operation is represented as $\mf{Q}\mf{K}^{\T} \oslash \mf{M}$, where $\oslash$ is the element-wise division operator and $\mf{M} \in \mb{R}^{N\times N}$ is the matrix containing  the norms of embedded features, with each element being:
\begin{equation}
    \mf{M}(n_1, n_2) = \|\mf{q}_{n_1}\|_{2}\|\mf{k}_{n_2}\|_{2}.
\end{equation}
Here, $\mf{q}_{n_1}$ denotes the $n_1$-$th$ row of $\mathbf Q$ and $\mf{k}_{n_2}$ denotes the $n_2$-$th$ row of $\mathbf K$. The scaling matrix  $\mf{M}$ is used to normalize the dot products, addressing the issue of energy imbalance caused by different missing patterns across extracted cubes. Then, the results are passed through the SparseMax function \cite{SparseMax}. This process generates the {\it sparse attention map}, { which is similar to a correlation matrix} and  is denoted as 
\begin{equation}
    \mf{S} = \text{SparseMax}(\mf{Q}\mf{K}^{\T}\oslash \mf{M}) \in \mb{R}^{N \times N}.
    \label{eq:create_s}
\end{equation}
The SparseMax function, like the SoftMax function, aims to produce a normalized score vector, but its output is much sparser (see illustrations in Fig.~\ref{fig:am} and brief implementation details in Appendix~C). Concretely, it selects a certain number of leading entries of the input while setting the rest to zero, and finally normalizes these support entries to sum up to 1. Further details can be found in \cite{SparseMax}.   

One might concern whether the missing values in $\bc Y$ could significantly affect the assessment of tensor attention.  However, our observation is that they do not. The reason is that when calculating the correlation between two high-dimensional signals, a few missing values in each signal will not substantially degrade the correlation estimation.


\subsubsection{Decoder Design}
\begin{figure}[t]
	\centering
	\includegraphics[width=1.01\linewidth]{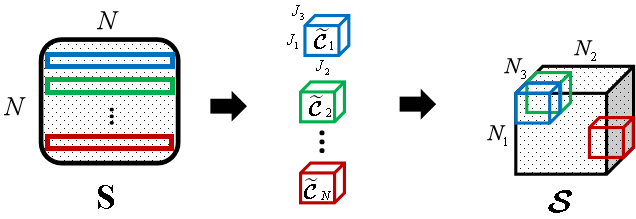}
	\caption{Illustration of  
 attention map tensorization.}
	\label{fig:tensorize}
\end{figure}
 Finally,  the  sparse attention map is used to reconstruct the entire tensor. This corresponds to what we call  ``decoder design'' in neural representation learning. 

The goal now is to tensorize the sparse attention map $\mf{S}$ into a sparse tensor $\bc{S}$ of appropriate dimensions,
which will act as the core tensor that interacts with the three factor matrices to produce the output. 
{ Note that $\bc{S}$ is the tensorized version of $\mf{S}$, containing  the same elements as $\mf{S}$.} To this end, the following procedure is adopted.

 The  sparse attention map 
 consists of $N$ row vectors (i.e, $\mf{S}=[\tilde{\mf{c}}_1, \cdots, \tilde{\mf{c}}_N]^{\T}$), with the $n$-$th$ row vector $\tilde{\mf{c}}_n^\T \in \mb{R}^{1 \times N}$ encoding the correlations between $n$-$th$ cube with other $N$ cubes. We first tensorize each row of $\mf{S}\in \mb{R}^{N\times N}$ into a sub-tensor (with size $J_1 \times J_2 \times J_3$). Specifically, the $n$-$th$ sub-tensor $\tilde{\bc{C}}_n$ can be represented as 
\begin{equation}
   \tilde{\bc{C}}_n(j_1, j_2, j_3) = \tilde{\mf{c}}_n((j_1-1)J_2J_3+(j_2-1)J_3+j_3).
\end{equation}
This process is done  orderly to ensure that the relative positions of the entries within each sub-tensor align with the positions of the corresponding cubes in $\bc{Y}$. Next,  we stack these sub-tensors in the same order to construct the sparse core tensor $\bc{S} \in \mb{R}^{N_1 \times N_2 \times N_3}$, where $N_l = (J_l)^2, \forall l$. That is,
\begin{equation}
   \bc{S}(\mf{n}_1, \mf{n}_2, \mf{n}_3) = \tilde{\bc{C}}_n,
\end{equation}
where the index set $\{\mf{n}_l = 1+(m_l-1)J_l:m_lJ_l, \forall l\}$ and  $m_1= (n-1) \| J_2J_3 + 1, m_2 = [(n-1) \mod J_2J_3] \| J_3 + 1, m_3 = (n-1) \mod J_3 +1$.  
The relative positions within each sub-tensor are also consistent with the arrangement of the extracted cubes in $\bc{Y}$.  As a result, the physical meaning of of an entry, 
 $\bc{S}(n_1,n_2,n_3)$, is that it quantifies  the correlation between the embedding feature of the $x$-$th$ and the $y$-$th$ cubes, where 
$ x = (n_1 \| J_1)J_2J_3 + (n_2 \| J_2)J_3 + (n_3 \| J_3)+1 $ and $ y = ((n_1 -1) \mod J_1)J_2J_3 + ((n_2 -1) \mod J_2)J_3 + ((n_3 -1) \mod J_3+1)$.
The tensorization process is illustrated in Fig.~\ref{fig:tensorize}, which can be simply understood as the reorganization of the elements of $\mathbf{S}$ into the tensor $\bc S$, thereby rendering $\bc{S}$ inherently sparse. Its primary goal is to preserve the spatial relationships of each entry, mirroring the relative positions of the cubes that are
extracted from $\bc{Y}$, see Fig.~\ref{fig:sa}. This ensures the construction of an informative core tensor that leverages the priors, which come from the observations, so that to effectively capture multidimensional interactions. The formulation of the tensorization process  can be conveniently implemented with PyTorch, as it presented in  Appendix~A.

The size of $\bc{S}$ determines the respective size of each one of the three learnable factor matrices (a.k.a value matrices)  $\{\mf{V}_1 \in \mathbb{R}^{I_1 \times N_1}, \mf{V}_2\in \mathbb{R}^{I_2 \times N_2}, \mf{V}_3\in \mathbb{R}^{I_3 \times N_3}\}$.
 The proposed sparse tensor attention module can be expressed as follows:
\begin{equation}
	\label{eq:sa}
	\begin{split}
		&\bc{G} = \text{STA}(\mf{Q}, \mf{K}, \mf{V}_1, \mf{V}_2, \mf{V}_3) =   \\
		& \bc{S} \times_1 \mf{V}_1 \times_2 \mf{V}_2 \times_3 \mf{V}_3  \in  \mathbb{R}^{I_1 \times I_2 \times I_3},\\
		&\text{s.t.}\quad \bc{S}=\text{Tensorize} (\text{SparseMax}(\mf{Q}\mf{K}^{\T}\oslash \mf{M}) ).
	\end{split}
\end{equation}

\begin{figure}[t]
	\centering
	\includegraphics[width=\linewidth]{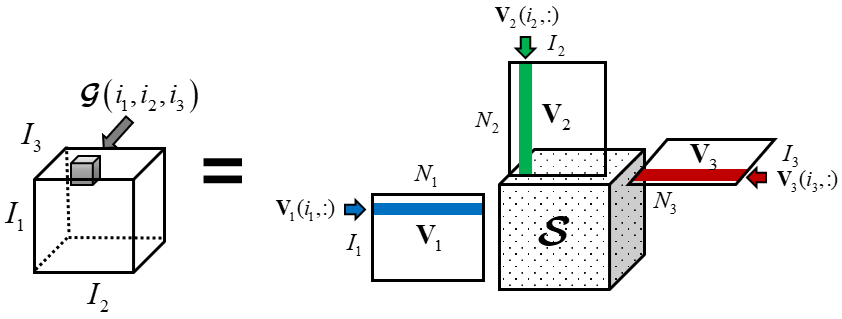}
	\caption{Element-wise view of the sparse tensor attention (STA) module.}
	\label{fig:ew}
\end{figure}

Essentially, the proposed sparse tensor attention (STA) module can be interpreted as the summation of all multiplicative interactions of the value matrices  weighted by the sparse core tensor $\bc{S}$. 
 More specifically, an element-wise interpretation of Eq.~\eqref{eq:sa} can be 
formulated as:
\begin{equation}
\bc{G}(i_1, i_2, i_3) = \bc{S} \times_1 \mf{V}_1(i_1, :) \times_2 \mf{V}_2(i_2, :) \times_3 \mf{V}_3(i_3, :).
\end{equation}
The illustration of this formulation can be seen in Fig.~\ref{fig:ew}. Note that $\mf{V}_{l}(i_l, :)$ can be interpreted as the  $i_l$-$th$ feature embedding of mode-$l$ (see the blue, green and red feature vectors in Fig.~\ref{fig:ew}). The sparse core tensor $\bc{S}$, which contains spatial  correlation weights, appropriately combines all the feature embeddings (i.e., $\mf{V}_{l}(i_l, :), \forall l$) and produces the output $\bc{G}(i_1, i_2, i_3)$.


The module operates intuitively: when the extracted cubes exhibit significant similarities, indicating a simpler representation model for the considered 3D physical field, the resulting sparse attention map highlights these similarities,  and it consequently leads to  a reduced number of the non-zero elements in the core tensor for reconstructing the 3D physical field, and vice versa. \textit{Therefore, the proposed module can adaptively adjust the model's complexity based on information that is extracted from the  the observations.}

\subsection{Proposed Tensor Attention Prior}
\label{s3}

In the previous two sections, we have discussed how to utilize a limited number of observations to construct a sparse attention map and then generate the core tensor for the Tucker model.  { The nonlinear activation function  $\varsigma(\cdot)$ is then introduced to produce the final output, aiming to further enhance the expressive power of the proposed model while maintaining stable gradients during training \cite{AF}. We recommend using Tanh or Sigmoid to regulate the output range.} The overall model, illustrated in Fig.~\ref{fig:sa}, is referred to as the \textit{tensor attention prior} (TAP) representation model. 


Given the TAP model, the problem of reconstructing the physical field can be formulated as follows:
\begin{equation}
	\label{crit}
	\begin{split}
		\min_{\mf{W}_{Q}, \mf{W}_{K}, \{\mf{V}_l\}_{l=1}^3 } &\left\|\bc{Y} - \bc{O} \ast  \varsigma(\bc{S} \times_1 \mf{V}_1 \times_2 \mf{V}_2 \times_3 \mf{V}_3) \right\|_{\F}^2, \\
		\text{s.t.} \quad &\bc{S}=\text{Tensorize} (\text{SparseMax}(\mf{Q}\mf{K}^{\T}\oslash \mf{M}) ),\\
		&\mf{Q}=\mf{P}\mf{W}_{Q}, \mf{K}=\mf{P}\mf{W}_{K}.
	\end{split}
\end{equation}

 Once the parameters  ${\mf{W^*}_{Q}, \mf{W^*}_{K}, \{\mf{V^*}_l\}_{l=1}^3 }$ are learned, the reconstructed field is given by:
\begin{align}
	\label{eq:output}
	\bc X = & \varsigma  (\text{Tensorize}  (\text{SparseMax} (\left[\mf{P}\mf{W^*}_{Q} \right] \left[ \mf{P}\mf{W^*}_{K} \right]^{\T} \oslash \mf{M}  ) \nonumber \\
	&~~~~~~~~~~~~~~~~~~~~~~~~~~  \times_1 \mf{V^*}_1 \times_2 \mf{V^*}_2 \times_3 \mf{V^*}_3   )  ).
\end{align}
The resulting 3D FieldFormer algorithm for 3D physical field reconstruction is presented in {\bf Algorithm~\ref{A1}}.

\begin{algorithm}[!t]
	\caption{3D FieldFormer based on TAP.} 
	\label{A1}
	{\bf Input:}
	Observations $\bc{Y} \in \mb{R}^{I_1\times I_2\times I_3}$, binary tensor $\bc{O} \in \mb{R}^{I_1\times I_2\times I_3}$, window size in three modes $(K_1, K_2, K_3)$, stride size in three modes $(S_1, S_2, S_3)$; \\
	{\bf Initialization:} Initialize the query and key matrices $\mf{W}_Q, \mf{W}_K \in \mb{R}^{M\times M}$ as well as the value matrices $\mf{V}_1 \in \mb{R}^{I_1\times N_1}, \mf{V}_2 \in \mb{R}^{I_2\times N_2}, \mf{V}_3 \in \mb{R}^{I_3\times N_3}$. 
	\begin{algorithmic}[1]
		\State Extract cubes from observations $\bc{Y}$ to get $\mf{P}$.
		\While{not converge} 
		\State Compute the query and key through  $\mf{Q}=\mf{P}\mf{W}_Q $, $\mf{K}=\mf{P}\mf{W}_K$.
		\State Compute the output of sparse attention module through Eq.~\eqref{eq:sa} and obtain reconstructed 3D physical field $\bc{X}$ via Eq.~\eqref{eq:output}.
		\State Compute the  loss $\|\bc{Y} -\bc{O}\ast\bc{X} \|_{\F}^{2}$
		\State Update $\mf{W}_Q, \mf{W}_K, \mf{V}_1, \mf{V}_2, \mf{V}_3$ according to the loss using  the Adam optimizer.
		\EndWhile
		\State \textbf{end while}
	\end{algorithmic}
	{\bf Output:}
	The reconstructed 3D physical field $\bc{X}$.
\end{algorithm}

\quad 

{ \textit{Remark 1 (Initializations and Convergence):} All the parameters are initialized using samples drawn from uniform distributions following the Kaiming initialization scheme \cite{kaiming}. We employ the Adam optimizer to minimize the objective function. Given that the objective is differentiable with respect to all parameters, the convergence of Adam to a stationary point over long-run iterations with a constant stepsize has been established in  \cite{adam}.}

\subsection{Multi-Head Tensor Attention Prior}
\label{s4}

\begin{figure}[t]
	\centering
	\includegraphics[width=1\linewidth]{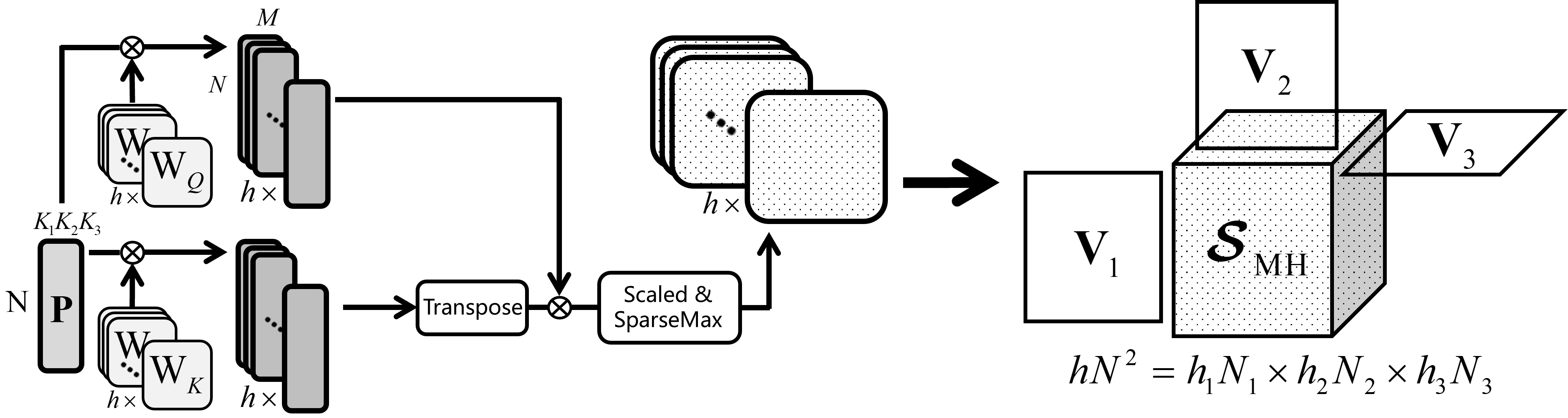}
	\caption{Illustration of Multi-head Sparse Tensor Attention Module.}
	\label{fig:mhsa}
\end{figure}
\vspace{-1mm}
To further enhance the  extracted information, the Multi-Head Sparse Tensor Attention (MHSTA) generalization is also proposed, where one can utilize multiple embedding subspaces for quantifying similarity and computing the corresponding attention maps. 
This is illustrated in Fig.~\ref{fig:mhsa}.
First,  $h$ query and key matrices are employed to perform, in parallel,   inner products followed by the corresponding  SparseMax operations.  Thus,  $h$ sparse attention maps are obtained.  
These maps  are then reshaped\footnote{We first tensorize $h$ attention maps   individually as before and obtain $\{\bc{S}_i \in \mathbb{R}^{N_1 \times N_2 \times N_3}\}_{i=1}^{h}$. Then, to make the sparse core tensor has the appropriate dimensionality (each dimension is relatively balanced), we choose $h_1, h_2, h_3$ such that $h=h_1h_2h_3$.  $\{\bc{S}_i \in \mathbb{R}^{N_1 \times N_2 \times N_3}\}_{i=1}^{h}$ are then stacked along the corresponding modes 
 to form $\bc{S}_{\text{MH}} \in \mathbb{R}^{h_1N_1 \times h_2N_2 \times h_3N_3} $ with multidimensional correlations preserved.} to obtain $\bc{S}_{\text{MH}} \in \mathbb{R}^{h_1N_1 \times h_2N_2 \times h_3N_3}$.  The Tucker model is applied next, where  $\bc{S}_{\text{MH}}$ operates with three sets of value matrices $\{\mf{V}_1 \in \mathbb{R}^{I_1 \times h_1N_1}, \mf{V}_2\in \mathbb{R}^{I_2 \times h_2N_2}, \mf{V}_3\in \mathbb{R}^{I_3 \times h_3N_3}\}$.  Note that equation $hN^2=h_1N_1 \times h_2N_2 \times h_3N_3$ should be guaranteed. This produces the output:
\begin{equation}
	\begin{split}
		\text{MHSTA}&(\{\mf{Q}_i\}_{i=1}^h, \{\mf{K}_i\}_{i=1}^h, \mf{V}_1, \mf{V}_2, \mf{V}_3) =\\
		&\bc{S}_{\text{MH}} \times_1 \mf{V}_1 \times_2 \mf{V}_2 \times_3 \mf{V}_3 \in  \mathbb{R}^{I_1 \times I_2 \times I_3}\\
		&\text{s.t.} \quad \bc{S}_{\text{MH}} =
		\text{Tensorize} (\text{SparseMax}(\\ &\text{Concat}(\mf{Q}_1\mf{K}_1^{\T} \oslash \mf{M}_1, \cdots, \mf{Q}_h\mf{K}_h^{\T}\oslash \mf{M}_h))),
	\end{split}
\end{equation}
where $\mf{Q}_i = \mf{P}\mf{W}_{Q}^{i}$, $\mf{K}_i = \mf{P}\mf{W}_{K}^{i}$ and $\mf{M}_i$ is computed from the row-wise norms of  $\mf{Q}_i$ and $\mf{K}_i$.

Then, we propose the \textit{multi-head tensor attention prior} (MHTAP) architecture by simply replacing the  STA of TAP with multi-head STA (MHSTA). The formulation of MHTAP for reconstructing 3D physical fields is as follows:	\begin{equation}
	\label{crit2}
	\begin{split}
		\min&_{\{\mf{W}_{Q}^i, \mf{W}_{K}^i\}_{i=1}^h, \{\mf{V}_l\}_{l=1}^3 } \\
		&\left\|\bc{Y} - \bc{O} \ast  \varsigma(\bc{S}_{\text{MH}} \times_1 \mf{V}_1 \times_2 \mf{V}_2 \times_3 \mf{V}_3) \right\|_{\F}^2.
	\end{split}
\end{equation}
The associated  FieldFormer algorithm is similar to {\bf Algorithm \ref{A1}} and summarized in  Appendix~B.

\quad

{ \textit{Remark 2 (Extension to Higher-order Field Tensors):} The proposed FieldFormer naturally extends to higher-order tensors by extracting multidimensional cubes, computing their similarities to generate a sparse attention map, and tensorizing it into a multidimensional core as described in Sec.~\ref{s2}. Finally, reconstruction is carried out using a multidimensional Tucker model.}

\section{Recoverability Analysis}
\label{sec:4}

In this section, we investigate the recoverability of the ground-truth 3D field tensor $\bc{X}_{\natural}$ based on the problem formulated in \eqref{crit}. Our analysis adapts the framework established in \cite{Radio, cn, ltr} to the proposed nonlinear Tucker model that includes a sparse core tensor with its sparsity pattern determined by attention schemes, non-orthogonal factor matrices, and a nonlinear activation function. Notably, we derive new theoretical results on the covering number and generalization errors associated with the proposed model.

To proceed,  we denote $\Omega$ as the set of observed indices, i.e., ${\Omega}=\{(i_1,i_2,i_3) | \bc{O}(i_1,i_2,i_3)=1\}$ and introduce the following definitions.

\textit{Definition 1 (Solution Set):} Let  $\mathcal{X}_{\text{TAP}}\subset \mb{R}^{I_1 \times I_2 \times I_3} $  be the {\it solution set} that contains all solutions $\tilde{\bc{X}}$ of \eqref{crit} satisfying  $\tilde{\bc{X}}=\varsigma(\tilde{\bc{S}} \times_1 \tilde{\mf{V}}_1 \times_2 \tilde{\mf{V}}_2 \times_3 \tilde{\mf{V}}_3) $, where  $\|\tilde{\bc{S}}\|_{\F} \le \alpha$ and $\|\tilde{\mf{V}}_l\|_{\F} \le \beta$ for $i = 1, 2, 3$.

\textit{Definition 2:} Define  $\text{Gap}(\tilde{\bc{X}}, \Omega)= \sqrt{\text{loss}_1(\tilde{\bc{X}})}-\sqrt{\text{loss}_2(\tilde{\bc{X}})}$, where   
\begin{equation}
	\begin{split}
		&\text{loss}_1(\tilde{\bc{X}}) = \frac{1}{\left |\Omega\right | } \sum_{(i_1,i_2,i_3) \in \Omega } \|\tilde{\bc{Y}}(i_1,i_2,i_3) - \tilde{\bc{X}}(i_1,i_2,i_3)\|_{2}^{2},\\
		&\text{loss}_2(\tilde{\bc{X}}) = \frac{1}{I_1I_2I_3 } \sum_{i_1,i_2,i_3 } \|\tilde{\bc{Y}}(i_1,i_2,i_3) - \tilde{\bc{X}}(i_1,i_2,i_3)\|_{2}^{2}.
	\end{split}
\end{equation}
Note that $\text{loss}_1(\tilde{\bc{X}})$ represents the loss measured on the observed part of the data, while  $\text{loss}_2(\tilde{\bc{X}})$ represents the loss measured over the entire data. Thus, $\text{Gap}(\tilde{\bc{X}}, \Omega)$ can be interpreted as the {\it generalization error}.

%
%
%

We can then show the following lemmas based on the definitions provided above.

\textit{Lemma 1:} The expected number of the non-zero elements in the sparse core tensor $\bc{S} \in \mb{R}^{N_1 \times N_2 \times N_3}$ is $\mb{E}(\|\bc{S}\|_{0}) = N\mb{E}(n(\mf z))$, where $n(\mf z)$ is a random variable representing the number of the non-zero elements in each row of the sparse attention map $\mf{S}$. The expectation of the non-zero elements $\mb{E}(n(\mf z))= \sum_{i=1}^{N}iP(n(\mf z) = i)$, where $P(n(\mathbf{z}))$  follows the distribution specified in (\ref{eq:probality}) under the assumption  that all entries of $\mf{Q}\mf{K}^{\T}\oslash \mf{M}$  are independent random variables drawn from a normal distribution with a mean of 0 and a variance of  1 \cite{attention}.
\begin{equation}
	\begin{split}
		\label{eq:probality}
		&P(n(\mf z)=1)=2-2\Phi(\frac{1}{\sqrt{2}}),\noindent \\ 
		&P(n(\mf z)=n)=(2-2\Phi(\frac{1}{\sqrt{2n}  })) \times \prod_{i=2}^{n}(2\Phi(\frac{1}{2i-2})-1),\noindent\\
		&~~~~~~~~~~~~~~~~~~~~~~~~~~~~~~~~~~~~~~~~~~~~n = 2,\cdots, N-1,\noindent\\
		&P(n(\mf z)=N)=\prod_{i=2}^{N}(2\Phi(\frac{1}{\sqrt{2i-2} })-1 ),
	\end{split}
 \vspace{-3mm}
\end{equation}
where $\Phi\left(\cdot \right)$ is the cumulative distribution function (CDF) of Gaussian distribution with mean 0 and variance 1:
\begin{equation}
	\Phi\left(X \right)= \int_{-\infty  }^{X} \frac{1}{\sqrt{2\pi}}e^{-\frac{x^2}{2}} dx.		
\end{equation} 

\textit{Proof}: See Appendix~D.\hfill  $\blacksquare$

Lemma 1 characterizes the expected number of the non-zero elements within the sparse tensor attention map, a metric useful for assessing the complexity of the proposed model. Building on the similar assumption made in the attention analysis using the SoftMax operator\cite{attention}, Lemma 1 establishes the sparsity level result for the SparseMax operator. This lemma is useful for future endeavors that employ the SparseMax operator to derive attention scores.

\textit{Lemma 2:} The covering number \cite{coveringnumber}  of the $\varepsilon$-net  of the solution set $\mathcal{X}_{\text{TAP}}$  is given by
\begin{equation}
	\label{eq:lemma2}
	\begin{split}
		&\mathsf N(\mathcal{X}_{\text{TAP} }, \varepsilon) \le \\
		&\left[\frac{3T(\beta^{3} +3\alpha\beta^{2})}{\varepsilon}\right]^{\|\bc{S}\|_{0}  +\sum_{l=1}^{3}N_lI_l }\alpha^{\|\bc{S}\|_{0}} \beta^{\sum_{l=1}^{3}N_lI_l},
	\end{split}
\end{equation}
where $T$ is the Lipschitz constant of the activation function  and $\mathbb{E}(\|\bc{S}\|_{0})$ is given by \textit{Lemma 1}.

\textit{Proof}: See Appendix~E.\hfill  $\blacksquare$

\textit{Lemma 3:} Let $\text{Gap}^{\ast}(\Omega)=\sup_{\tilde{\bc{X}} \in \mathcal{X}_{\text{TAP}} }\left |\text{Gap}(\tilde{\bc{X}}, \Omega) \right |$  be the supremum of $\text{Gap}(\tilde{\bc{X}}, \Omega)$. Based on the sensing model in (\ref{om}), the following inequality holds with a probability of at least $1-\delta$:
\begin{equation}
	\label{eq:lemma3}
	\text{Gap}^{\ast}(\Omega) \le \frac{2\varepsilon }{\sqrt{\left |\Omega\right |}} + \left(\frac{\xi^2\omega}{2}\log(\frac{2\mathsf N(\mathcal{X}_{\text{TAP} }, \varepsilon)}{\delta})  \right)^{\frac{1}{4}},
\end{equation} 
where $\varepsilon > 0$ is a positive scalar and $\omega = (\frac{1}{\left |\Omega\right | }+ \frac{1}{\left |\Omega\right |I_1I_2I_3  } -\frac{1}{I_1I_2I_3 })$. $\mathsf N(\mathcal{X}_{\text{TAP} }, \varepsilon)$ is the covering number of $\mathcal{X}_{\text{TAP} }$ (see \textit{Lemma 2}),  and $\xi = \left(\nu+v+\alpha\beta^3\right)^2$ with $\nu = \max_{i_1,i_2,i_3} \left|\bc{X}_{\natural}(i_1,i_2,i_3)  \right| $ and $v = \max_{i_1,i_2,i_3} \left|\bc{N}(i_1,i_2,i_3)  \right| $.

\textit{Proof}: See Appendix~F.\hfill  $\blacksquare$	

Lemma 2 and Lemma 3 characterize the generalization error for the proposed TAP model. They extend the results developed in \cite{Radio, cn, ltr} to account for the sparse tensor core, non-orthogonal factors, and the non-linear activation function. As a result, in addition to paving the way for establishing the recoverability of the proposed model (refer to Theorem 1 below), these two lemmas themselves are essential for quantifying the generalization error in any future nonlinear tensor Tucker model.

Finally, we can derive the main recoverability theorem based on the aforementioned lemmas and definitions.

\textit{Theorem 1(Recoverability):}
Under the sensing model described in (\ref{om}), assume that  $\bc{X}^{\ast} =\varsigma(\bc{S}^{\ast} \times_1 \mf{V}^{\ast}_1 \times_2 \mf{V}^{\ast}_2 \times_3 \mf{V}^{\ast}_3)$, where $\bc{S}^{\ast}$ and $\{\mf{V}^{\ast}_l \}_{l=1}^3$  are obtained from any optimal solution of \eqref{crit}. Here, $\bc{S}^{\ast}$ represents a sparse core tensor and the activation function  $\varsigma(\cdot)$ is $T$-Lipschitz continuous. Also assume that $\bc{S}^{\ast}$ and $\{\mf{V}^{\ast}_l \}_{l=1}^3$ satisfy the specifications in the solution set in \textit{Definition 1}.  Then, with a probability of at least $1-\delta$, the following statement holds:
\begin{equation}
	\label{recoverbility}
	\begin{split}
		&\frac{\|\bc{X}^{\ast} - \bc{X}_{\natural} \|_{\F}}{\sqrt{I_1I_2I_3}}  \le \text{Gap}^{\ast}(\Omega) + (\frac{1}{\sqrt{I_1I_2I_3}}\left\| \bc{N}\right\|_{\F}+ \\ &\frac{1}{\sqrt{\left | \Omega \right | }} \left\|\bc{O} \ast \bc{N} \right\|_{\F}) + \frac{1}{\sqrt{\left | \Omega \right | }}\left\| \tilde{\bc{X}}^{\ast} - \bc{X}_{\natural} \right\|_{\F}.
	\end{split}
\end{equation}
where $\tilde{\bc{X}}^{\ast} = \arg\min_{\tilde{\bc{X}} \in \mathcal{X}_{\text{TAP}}} \left\|\tilde{\bc{X}}- \bc{X}_{\natural}  \right\|_{\F}^2$ and $\text{Gap}^{\ast}(\Omega)$ is upper bounded by \eqref{eq:lemma3} with the covering number in \eqref{eq:lemma2}.

\textit{Proof}: See Appendix~G.\hfill  $\blacksquare$	


Theorem 1 presents a bound on the estimation error for recovering the ground-truth 3D physical fields,  based on the criterion described  in (\ref{crit}). Upon closer inspection of \eqref{recoverbility}, it is readily seen that the  recovery error  stems from three primary sources: the  generalization error $\text{Gap}^{\ast}(\Omega)$, the sensing noise $\bc{N}$, and the representation error of the TAP model. Consequenlty, exact recovery occurs only when these three error sources simultaneously approach zero. This necessitates the noise to diminish, and the  representation model to strike an optimal balance between conciseness (resulting in near-zero generalization error) and expressiveness (resulting in near-zero representation error).

Note that the number of the non-zero elements of $\bc{S}$ represents a trade-off between the conciseness and expressiveness of TAP model, thereby influencing the generalization error $\text{Gap}^{\ast}(\Omega)$. An increase in the number of the non-zero elements in the attention map amplifies the generalization error $\text{Gap}^{\ast}(\Omega)$, but at the same time it enhances the model's capacity to represent a more intricate 3D field, thereby reducing the representation error $\left\| \tilde{\bc{X}}^{\ast} - \bc{X}_{\natural} \right\|_{\F}$,  and vise versa.

 { Further discussions, concerning comparisons  of the recoverability analysis with the existing model \cite{Radio}, can be found in Appendix~H. 
 	 The impact of the sparsity level and observation patterns on the recoverability analysis is discussed in Appendix~M.  The main insight is that the proposed model has the potential to achieve lower reconstruction error than \cite{Radio}, particularly under distribution shifts.
This advantage stems from the self-supervised learning of $||\bc{S}||_{0}$, which allows the model's covering number (see Lemma 2) to adapt to the observed data, thereby reducing the generalization error.
Moreover, since our model does not rely on any pretrained models, its representation error remains unaffected by distribution shifts. }Our recoverability analysis above can also be easily extended to the MHTAP model.

\section{Experiments and Discussions}
\label{sec:5}
In this section, we present experimental results to demonstrate the effectiveness and versatility of the proposed methods using two 3D physical field datasets. We first compare our methods against state-of-the-art (SOTA) techniques and then carry out the ablation studies to draw insights concerning the proposed approach. The corresponding algorithms  are implemented using PyTorch 1.13.1 and all experiments are performed on a RTX 4070 GPU with 8 GB of GPU memory.	
\subsection{Ocean Sound Speed Field (SSF) Reconstruction}
{\it 3D SSF data:}  In the following experiments, the South China Sea SSF dataset denoted as $\bc{X}_{\natural} \in \mb{R}^{20 \times 20 \times 20}$ is utilized \cite{TDL,TNN}. The dataset covers a spatial area of $152\text{km} \times 152\text{km} \times 190 \text{m}$ and has a horizontal resolution of 8 km and a vertical resolution of 10 m.

{\it Performance metric:}  	The reconstruction performance is evaluated by the root mean square error (RMSE) \cite{TNN}:
\begin{equation}
	\text{RMSE} = \sqrt{\frac{1}{I}\|\bc{X}-\bc{X}_{\natural} \|_{\F}^{2} },
\end{equation}
where $\bc{X}$ and $\bc{X}_{\natural}$ represent the reconstructed SSF and the ground truth, respectively. The total number of SSF entries  $I $ equals  $ I_1 \times I_2 \times I_3$. The reported RMSEs are averaged over 10 Monte-Carlo trials.

{\it Baseline:} Three unsupervised reconstruction methods, namely Tucker-ALS \cite{Tucker-als}, LRTC \cite{LRTC}, and TNN \cite{TNN} are selected.  The Tucker-ALS and LRTC methods are model-based approaches that utilize handcrafted priors, specifically the low-rankness property. The TNN method, on the other hand, can be regarded as an untrained deep-learning version of the Tucker-ALS method.

{\it Implementation Details:} See Appendix~I.

\begin{table}[!t]
	\centering
	\caption{Average reconstruction errors (RMSEs) of the proposed methods and the benchmarks for different $\rho$ values. The bold and underlined numbers represent the lowest and second lowest RMSEs in the comparisons, respectively.}
	\renewcommand{\arraystretch}{1.5}
	\begin{tabularx}{0.9\linewidth}{l c c c c }
		\toprule
		Methods&  $\rho=5\%$ &$\rho=10\%$ &$\rho=20\%$ & $\rho=30\%$     \\	
		\midrule
		Tucker-ALS & 2.666& 1.527& 0.663 & 0.411  \\
		LRTC & 2.723 & 2.228 & 1.181 & 0.773  \\
		TNN & 2.803 & 1.562  & 0.405 & 0.312   \\
		TAP & \textbf{0.954} & \underline{0.560} & \underline{0.346} & \underline{0.245} \\
		MHTAP & \underline{1.225} & \textbf{0.535} & \textbf{0.317} &  \textbf{0.218}  \\
		\bottomrule
		\label{Tab:1}
	\end{tabularx}
 \vspace{-5mm}
\end{table}


\begin{figure*}[t]
	
	\begin{minipage}{\textwidth}
 \centering
\includegraphics[width=0.85\linewidth]{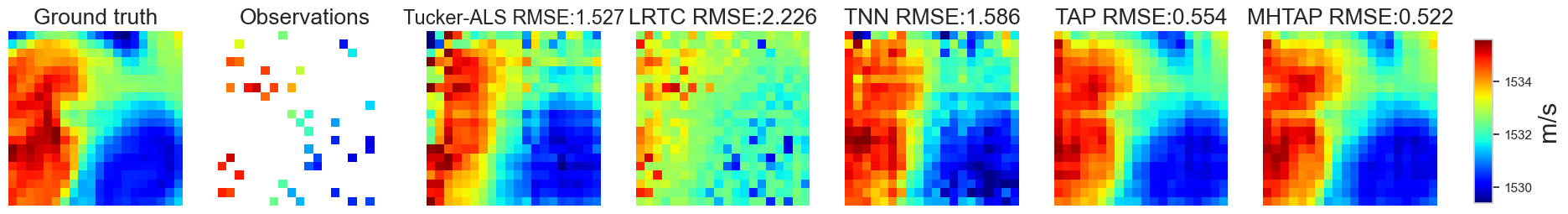} 
	\end{minipage}
 \begin{minipage}{\textwidth}
 \centering
		\includegraphics[width=0.84\linewidth]{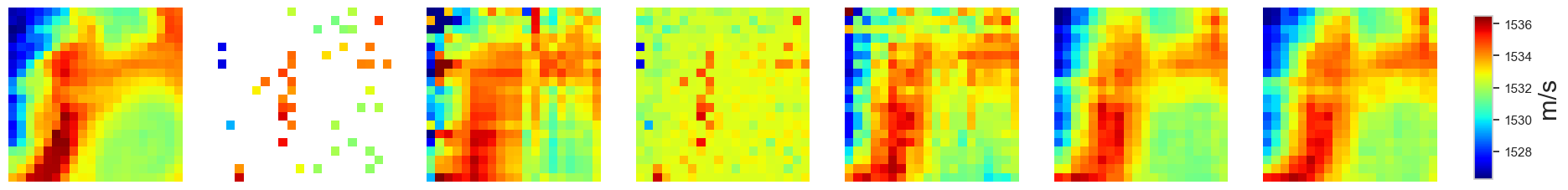} 
	\end{minipage}
  \vspace{-0.1cm}
  \begin{minipage}{\textwidth}
  \centering
		\includegraphics[width=0.82\linewidth]{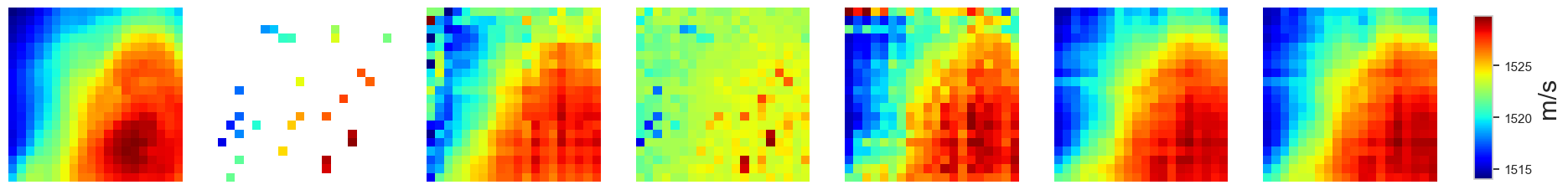} 
	\end{minipage}
 \caption{{ 
Visualizations of ground-truth, observation and reconstructed SSF data of various methods at depth 0m (top), 100m (middle), 200m (bottom) under  $\rho = 10\%$. The corresponding RMSEs are provided at the top of the figure.
 }}
\label{fig:ssf_visual}
\end{figure*}


{\it Results:}  We first test the proposed methods using the noise-free SSF data under various observation rates  $\rho=\frac{\|\bc{O}\|_{0}}{I_1\times I_2 \times I_3}$.

Table~\ref{Tab:1} presents the RMSEs of different algorithms across different $\rho$ values. As the observation rate increases, the reconstruction errors decrease for all algorithms. Notably, when $\rho \ge 20\%$, TNN demonstrates significant advantages over the model-based methods (Tucker-ALS and LRTC). Furthermore, the proposed methods consistently outperform the SOTA baselines in all scenarios. MHTAP exhibits superior performance compared to TAP when $\rho$ is higher ($\rho \ge 10\%$). This is attributed to the MHTAP's ability to capture more information from different subspaces of the input as more measurements of the SSF are observed. However, when $\rho$ is relatively low, MHTAP tends to overfit the limited observations due to its increased number of parameters. Fig.~\ref{fig:ssf_visual} illustrates the reconstructed SSF data of the five methods and their corresponding RMSEs at $\rho=10\%$. TAP and MHTAP provide more accurate fits to the missing entries. Overall, the proposed methods achieve significantly superior results in both quantitative and visual evaluations compared to the very recent TNN model.

Next, we evaluate the performance of the proposed methods using noisy SSF data, where Gaussian noise with a mean of 0 and variance  $\sigma^2$ is added according to the sensing model \eqref{om}. Table~\ref{Tab:noise} presents the RMSEs of different algorithms at various sampling ratios and noise powers. It is observed that the TNN algorithm proves ineffective at a low observation rate ($\rho =10\%$). In contrast, the proposed methods outperform all the baseline algorithms across all  $\rho$ values. Particularly, when the observation rate is very low (e.g., $\rho =10\%$), TAP exhibits superior noise robustness compared to MHTAP, primarily due to its reduced number of parameters. Conversely, when the observation rate is high ($\rho \ge 20\%$), MHTAP achieves better performances than TAP.		

{ To evaluate the robustness of our method, we add to the observations  Laplacian noise with a mean of 0 and variance $\sigma^2$  and assess the reconstruction performances. As shown in Table \ref{Tab:noise2}, the reconstruction error increases only slightly in the presence of Laplacian noise,  compared to Gaussian noise of the same power, indicating the robustness of our method to non-Gaussian noise sources.}

\begin{table}[t]
	\centering
	\caption{Average reconstruction errors (RMSEs) of the proposed methods and the benchmarks for different $\rho$ values and  Gaussian noise powers. The bold and underlined numbers represent the lowest and second lowest RMSEs in the comparisons, respectively.}
	\renewcommand{\arraystretch}{1.5}
	\begin{tabularx}{0.9\linewidth}{c l  c c c }
		\toprule
		& Methods &$\rho=10\%$ & $\rho=20\%$  & $\rho=30\%$   \\	
		\midrule
		& Tucker-ALS &  1.548 &0.655 & 0.412 \\
		& LRTC &  2.232 &1.195 & 0.789 \\
		$\sigma=0.1$&TNN  &1.683&0.471 & 0.329  \\
		&TAP  &\textbf{0.571}& \underline{0.382} & \underline{0.293} \\
		&MHTAP  &\underline{0.609}&\textbf{0.346} & \textbf{0.261} \\
		\midrule
		& Tucker-ALS&1.546 &0.663 & 0.425 \\
		& LRTC &  2.247 &1.226 & 0.826 \\
		$\sigma=0.2$&TNN  &1.784&0.516 & 0.356  \\
		&TAP  &\textbf{0.600}& \underline{0.425} &\underline{0.351} \\
		&MHTAP  &\underline{0.610}&\textbf{0.406} & \textbf{0.318} \\
		\midrule
		& Tucker-ALS&1.549 &0.639 & 0.451 \\
		& LRTC &  2.266 &1.270 & 0.875 \\
		$\sigma=0.3$&TNN  &2.170&0.561 & 0.423  \\
		&TAP  &\textbf{0.685}& \underline{0.482} & \underline{0.419} \\
		&MHTAP  &\underline{0.714}&\textbf{0.471} & \textbf{0.388} \\
		\bottomrule
		\label{Tab:noise}
	\end{tabularx}
  \vspace{-5mm}
\end{table}

\begin{table}[t]
	\centering
	\caption{Average reconstruction errors (RMSEs) of the proposed methods for different $\rho$ values and Laplacian noise powers.}
	\renewcommand{\arraystretch}{1.5}
	\begin{tabularx}{\linewidth}{c l  c c c }
		\toprule 
		& { Methods} &{$\rho=10\%$} & {$\rho=20\%$}  & {$\rho=30\%$}   \\	
		\midrule
		{$\sigma=0.1$}&{TAP (Laplacian)} &{0.617} &{0.385}
		&{0.302}   \\
		&{MHTAP (Laplacian)}  &{0.633} &{0.356} &{0.283} \\
		\midrule
		{$\sigma=0.2$}&{TAP (Laplacian)}  &{0.622} &{0.424}  & {0.351}  \\
		&{MHTAP (Laplacian)}  &{0.662} & {0.410} &{0.336} \\
		\midrule
	{$\sigma=0.3$}&{TAP (Laplacian) } &{0.717} & {0.487}  & {0.426}  \\
		&{MHTAP (Laplacian)}  &{0.745} &{0.491}  &{0.420} \\
		\bottomrule
		\label{Tab:noise2}
	\end{tabularx}
  \vspace{-5mm}
\end{table}

\subsection{Radio Map Tensor  Reconstruction}

\begin{figure*}[t]
	
	\begin{minipage}{\textwidth}
 \centering
\includegraphics[width=0.9\linewidth]{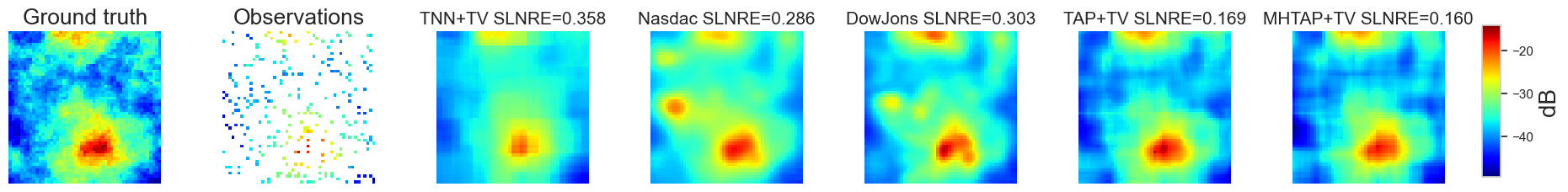} 
	\end{minipage}
 \begin{minipage}{\textwidth}
 \centering
		\includegraphics[width=0.9\linewidth]{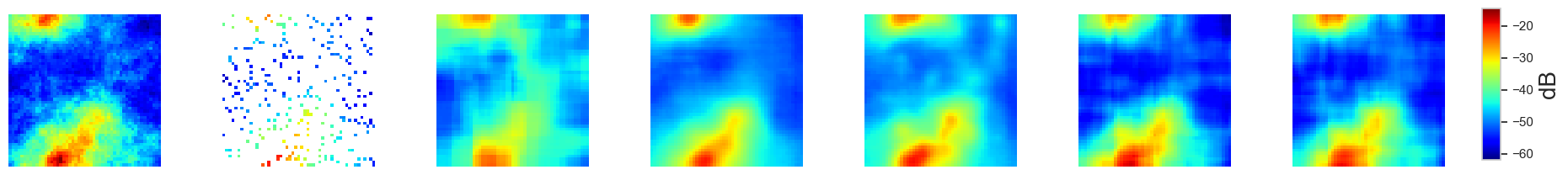} 
	\end{minipage}
  \vspace{-0.1cm}
  \begin{minipage}{\textwidth}
  \centering
		\includegraphics[width=0.9\linewidth]{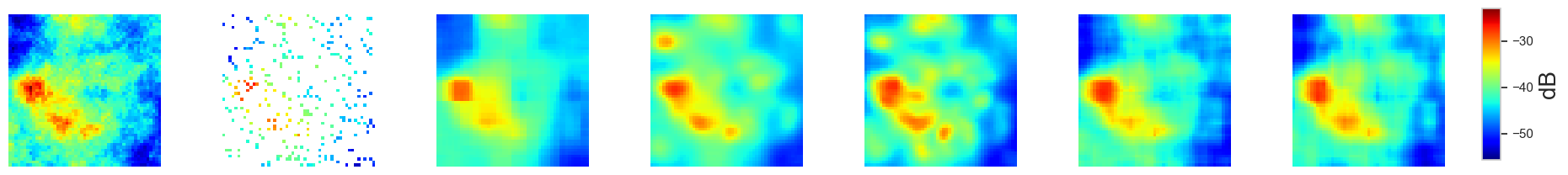} 
	\end{minipage}
 \caption{Visualizations of ground-truth, observation and reconstructed radio maps of various methods at the 7-$th$ (top), 18-$th$ (middle) and 29-$th$ (bottom) frequency bins; $\rho = 10\%$, $R = 7$, $d_{\text{corr}} = 50$, $\eta = 10$. The corresponding SLNREs are provided at the top of the figure.  }
\label{fig:rm_visual}
\end{figure*}

		\begin{figure*}[t]
	\centering
	\includegraphics[width=0.95\linewidth]{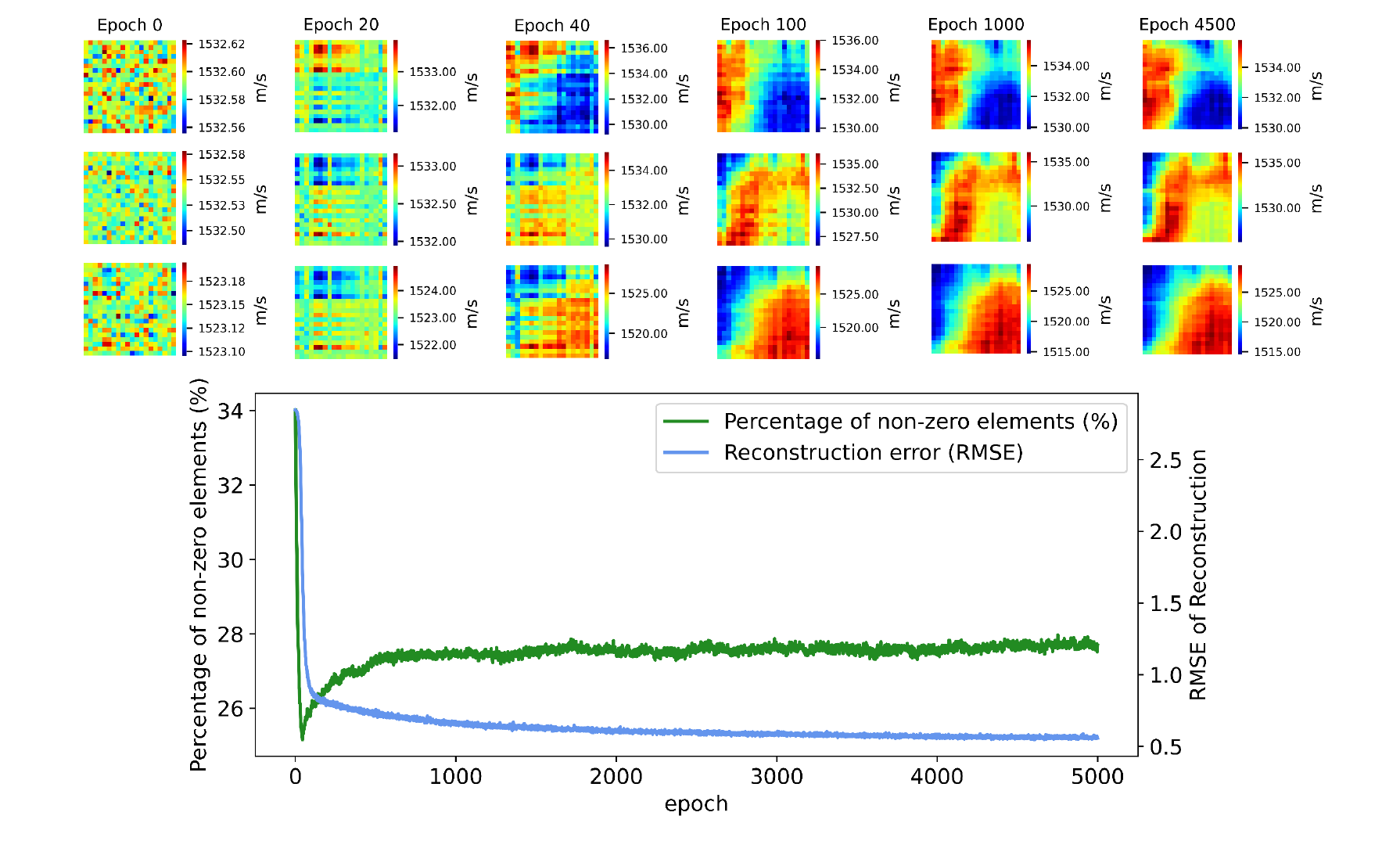}
	\caption{ Reconstruction results of the proposed FieldFormer  on  SSF data with $10\%$ observations over epochs. \textbf{Top}: Visualizations of reconstruction results over various epochs on depth 0m (top), 100m (middle), 200m (bottom). \textbf{Bottom}: Illustration of the percentage of the non-zero elements in the Tucker core tensor and the reconstruction error over epochs. }
	\label{fig:sparsityloss}
\end{figure*}
\vspace{-1mm}
In this section,  we conduct experiments on radio map tensor reconstruction.

{\it Data Description:}
The radio map tensors used in the following experiments	are generated from the joint path loss and log-normal shadowing model\cite{Radio,WC}, which can be expressed as:
\begin{equation}
	\label{rm_sm}
	\bc{X}_{\natural} = \sum_{r=1}^{R} \mf{S}_r \circ \mf{c}_r \in \mathbb{R}^{I_1 \times I_2 \times I_3}.
\end{equation}
Here, $R$ is the number of emitters  and  $\mf{S}_r\in \mathbb{R}^{I_1 \times I_2}, \mf{c}_r\in \mathbb{R}^{I_3}$ is the spatial loss function (SLF) and power spectrum density (PSD) of $r$-$th$ emitter. 
Specifically, the $r$-$th$ SLF   at location $\mf{m}=[i,j]^{\T}$  is generated by
\begin{equation}
	\mf{S}_r(i,j) = \|\mf{m}-\mf{m}_r\|_{2}^{-\gamma_r}10^{\frac{v_r(\mf{m})}{10}},
\end{equation}
where $\mf{m}_r$ is the coordinate of  $r$-$th$ emitter, $\gamma_r$ is the  $r$-$th$ path loss coefficient, and $v_r(\mf{m})$ is  the correlated  shadowing component  that is generated from Gaussian distribution with zero mean and variance $\eta_r$. The auto-correlation function between $\mf m$ and $\mf m^{'}$ is $\mb{E}(v_r(\mf{m}), v_r(\mf{m^{'})}) = \eta_r\exp(-\|\mf{m}-\mf{m}^{'}\|_2/d_{\text{corr}}) $, in which   $d_{\text{corr}}$  represents the decorrelation distance. The PSD of the emitter $r$ is 
given by $\mf{c}_r(k) = \sum_{i=1}^{M} a_{(i,r)} \text{sinc}^2 (k-f_{(i,r)}/w_{(i,r)})$, where $a_{(i,r)}$ is the scaling factor of $r$-$th$ emitter at $i$-$th$ subband drawing from a uniform distribution over the interval $(0.5,2.5)$; $f_{(1,r)}, \cdots, f_{(M,r)}$ are the central frequencies of $M$ subbands available to $r$-$th$ emitter and $M$ is set to 10. $w_{(i,r)}$ controls the  width of sidelobe of  $r$-$th$ emitter at $i$-$th$ subband drawing from an uniform distribution over the interval $(2, 4)$. In the experiments, $I_1 = I_2 = 51$, $I_3 = 64$.

{\it Performance metric:}
The performance metric is the scaled log-domain normalized reconstruction error (SLNRE) \cite{quant}:
\begin{equation}
	\text{SLNRE}=100 \times\frac{\| 
		\bc{X}_{\text{log}}-\bc{X}_{\natural\text{log}}\|_{\F}^2}{\|\bc{X}_{\natural\text{log}}\|_{\F}^2 },
\end{equation}
where $\bc{X}_{\text{log}}$ is the log-transformed reconstructed radio map while $\bc{X}_{\natural\text{log}}$ being the log-transformed  ground truth radio map. The SLNRE is an appropriate metric for skewed data like radio map \cite{quant}.  All the SLNREs are averaged over 10 Monte-Carlo trials. 

{\it Baseline:}
We use  SOTA untrained method TNN\cite{TNN} and trained methods Nasdac and  DowJons\cite{Radio}, to benchmark the proposed methods. Nasdac learns a generative deep network to reconstruct individual SLFs of the emitters from sensor measurements based on block term decomposition model. And DowJons leverages the generative deep network as structural constraints and reconstruct radio map tensors based on optimization criterion.	

{\it Implementation Details:} See Appendix~J.

{\it Results:} The first dataset (radio map 1) is generated using the typical setting described in \cite{Radio}, where  the number of emitters  $R=7$, the decorrelation distance  $d_{\text{corr}}=50$ and the shadowing coefficient  $\eta =6$   for all $r$. 
In the subsequent experiments, we use a fiber-wise sampling methodology to obtain observed radio map tensors, with sensors randomly  distributed across a 2D spatial domain to capture measurements over the entire frequency spectrum.

Table \ref{Tab:rm1} shows the SLNREs for the trained methods (Nasdac, DowJons), the untrained methods (TNN, TAP, MHTAP) and these untrained methods enhanced by total variation (TV) regularization \cite{TV} under various $\rho$ values on reconstructing radio map 1.  { Total variation  regularization provides complementary structual regularization on the smoothness of the 3D field and serves as a widely used, easily implemented handcrafted prior in reconstruction tasks.}
One can see that the proposed TAP and MHTAP consistently perform better than TNN, with MHTAP  consistently outperforming TAP. 
However, these untrained methods (TNN, TAP, MHTAP) are still inferior to Nasdac and DowJons, especially when $\rho$ is small, due to the lack of supervised prior knowledge. 
The performance gap can be simply filled in via incorporating a  simple TV regularization to enforce the spatial smoothness of the reconstructed maps  (details can be found in Appendix~K).  Specifically, the SLNRE of TNN decreases rapidly as $\rho$ increases and  surpasses Nasdac and DowJons when $\rho \ge 15\%$, indicating that TNN becomes effective given enough observations.  the proposed methods outperform all the baselines.  
\begin{table}[t]
	\centering
	\caption{Average reconstruction error (SLNREs) of different methods in reconstructing radio map 1  versus various $\rho$ values. Bold number and underlining number indicate the lowest and  the second lowest  SLNREs in the comparisons, respectively.  }				\renewcommand{\arraystretch}{1.5}
	\begin{tabularx}{0.9\linewidth}{l c c c c }
		\toprule
		Methods& $\rho=5\%$& $\rho=10\%$ & $\rho$=$15\%$ &  $\rho=20\%$      \\	
		\midrule
		Nasdac& 0.246&0.192 &0.179 & 0.161 \\
		DowJons & 0.216 &0.173 &0.167 & 0.149\\
		\midrule
		TNN &  83.65 & 25.72 & 1.574&0.337  \\
		TAP& 7.132 &0.914 &0.489 & 0.223\\
		MHTAP & 5.517 & 0.480 & 0.326 & 0.168\\
		\midrule
		
		TNN+TV &  0.539 &0.242 & 0.158 & 0.116 \\
		TAP+TV& \underline{0.189} &\underline{0.100} &\underline{0.073} & \underline{0.067}\\
		MHTAP+TV & \textbf{0.180} &\textbf{0.091} &\textbf{0.071} &\textbf{0.065}\\
		\bottomrule
		\label{Tab:rm1}
	\end{tabularx}
 \vspace{-5mm}
\end{table}

\begin{table}[t]
	\centering
	\caption{Average reconstruction error (SLNREs) of different methods in radio map reconstruction task under $\rho = 10\%$ versus various $\eta_s$. Bold number and underlining number indicate the lowest and  the second lowest  SLNREs in the comparisons, respectively.   } 
	\renewcommand{\arraystretch}{1.5}
	\begin{tabularx}{0.7\linewidth}{l c c c  }
		\toprule
		Methods& $\eta = 9$ & $\eta = 10$ &   $\eta = 12$      \\	
		\midrule
		Nasdac& 0.238 &0.286 &0.667  \\
		DowJons & 0.232 &0.303 & 0.631  \\
		\midrule
		TNN+TV & 0.304 &  0.369 & 0.411  \\
		TAP+TV& \underline{0.130} &\underline{0.170} & \underline{0.232} \\
		MHTAP+TV & \textbf{0.128} &\textbf{0.162} &\textbf{0.222} \\
		\bottomrule
		\label{rm2}
	\end{tabularx}
  \vspace{-8mm}
\end{table}

Fig.~\ref{fig:rm_visual} provides the illustrative example given  $\rho = 10\%, R=7, d_{\text{corr}}=50$ and $\eta=10$, bridging the SLNREs and the visual quality of the reconstructions. First, both Nasdac and Dowjons accurately reconstruct the radio map. Then, these untrained methods, further aided by TV regularization which enforces multidimensional smoothness of the outputs, achieve  good visual results given few observations. Finally, the proposed methods capture more fine-grained features of the radio map compared to the Nasdac and the DowJons.

Next, we consider  radio map tensors with  more complex shadowing environments.  We create the radio map tensors with   $R=10$, $d_{\text{corr}}=50$ under various $\eta_s$ (i.e., 9, 10, 12) for all $r$, which are the out-of-training distribution data for the Nasdac and the DowJons. 
In the subsequent experiments, the untrained methods are all enhanced by  TV regularization.

Table.~\ref{rm2} displays the  SLNREs of different methods given $\rho = 10\%$. 
One can see that  Nasdac and the DowJons deteriorate in reconstructing the radio maps that they have never seen before, being surpassed by TNN when $\eta = 12$.  With TV regularization, the TAP and MHTAP consistently outperform  all the baselines and  
MHTAP still remains  superior to TAP. These experiments verify that the proposed methods generalize better in handling 3D physical fields of varying complexity levels without the cost of extensive training and the risk of inaccurately estimating  the number of emitters.

{ We  then provide  detailed analysis of why the proposed approach outperforms the baseline methods in certain scenarios (e.g., at low observation rates), particularly by examining the specific properties of the proposed method. Both Tucker-ALS and TNN, like the proposed FieldFormer, are based on the Tucker model. However, they share a key limitation: the need to predefine the size of the Tucker core, which directly determines the model's complexity. Although we carefully tune hyperparameters (e.g., the multilinear rank) over a large search space to achieve their best overall performance, these methods -- due to their fixed, non-adaptive model complexity -- struggle to adapt to different observation scenarios, particularly at low observation rates. As shown in Fig.~\ref{fig:ssf_visual}, both methods tend to overfit the noise in sparse observations.
	
	 LRTC, which is based on the t-SVD model, seeks to minimize the tensor tubal rank while reconstructing the entire field from sparse observations. However, this objective function prevents LRTC from capturing the fine details of the physical field, leading to underfitting -- particularly when reconstructing complex structures such as the SSF, as it is illustrated in Fig.~\ref{fig:ssf_visual}. 
	
	 Nasdac and DowJons, which are trained on pre-collected data, rely on the assumption that the training and test data share similar distributions. However, in real-world scenarios, where data distribution shifts occur, these pre-trained models fail to generalize effectively, leading to significant performance degradation.
	
    In contrast, the proposed FieldFormer dynamically adjusts its model complexity to fit the observations without requiring training data. This adaptability helps mitigate {\it overfitting, underfitting, and sensitivity to data distribution shifts}, leading to superior performance compared to the baseline methods.}

\subsection{ Complexity-adaptive Neural Representation}
{In this subsection,  we provide detailed discussions and experimental results to explain how the attention mechanism adjusts model complexity. The proposed  FieldFormer, based on the Tucker model, leverages an attention mechanism to dynamically control the sparsity of the core tensor during the reconstruction process.  Fig.~\ref{fig:sparsityloss} provides detailed visualizations of the intermediate steps on reconstructing the SSF data. The top of  Fig.~\ref{fig:sparsityloss} illustrates reconstruction results at different epochs, while the bottom shows the percentage of the non-zero elements in the Tucker core tensor and the corresponding reconstruction error over epochs.  It is readily observed that the reconstruction results progress {\it from coarse to fine as the number of epochs increases}. The percentage of the non-zero elements drops sharply  before gradually rising to a plateau, while the reconstruction error consistently decreases. This phenomenon suggests that the model complexity is {\it dynamically adjusted} throughout the reconstruction process. 

At the beginning of reconstruction, the parameters of the attention mechanism (i.e., $\mf{W}_Q, \mf{W}_K$) are randomly initialized, causing similarity scores between different cubes to be randomly distributed. Consequently, the percentage of the non-zero elements in the core tensor is relatively high. As the model optimizes the loss function, the attention mechanism becomes more effective at capturing meaningful similarities between different cubes. This leads to a reduction in the percentage of the non-zero elements, as similarity scores concentrate on the most relevant pairs. More specifically, at the early epochs, since the parameters are updated in the direction of {\it the steepest descent} of the loss function,  the model primarily fits the lower-frequency components (e.g., the mean value) of the field, as shown in Fig.~\ref{fig:sparsityloss}. This is reflected in the rapid decrease in reconstruction error and the relatively coarse visual appearance of the reconstruction results. During this phase, model complexity remains low, as indicated by a low percentage of the non-zero elements in the Tucker core tensor. As training progresses, in order to further decrease the reconstruction loss, the model gradually captures higher-frequency components of the field, leading to an increase in model complexity. As shown in Fig.~\ref{fig:sparsityloss}, the reconstruction error declines more slowly in later stages, while the reconstruction results become increasingly fine-grained. This explains the subsequent rise in the percentage of the non-zero elements in the core tensor. Eventually, the model reaches an {\it optimal level of complexity} for representing the field, after which the percentage of the non-zero elements stabilizes. We further provide discussions on  the trade-off between computational cost and reconstruction accuracy in Appendix~N.

}

\subsection{Ablation Study}

\subsubsection{Impact of Prior Information from Partial Observations Incorporated by Attention Mechanism}

In this paper, we propose using dot-product  between query and key (generated from extracted cubes $\mf{P}$)  to gain informative priors from the limited observations.  This approach allows for adaptively tuning the sparsity of the core tensor in a self-supervised manner.

Therein, we conduct the ablation study on TAP/MHTAP in reconstructing radio map 1 to prove its efficacy. For comparison, we replace  $\mf P$ containing observation information with  noise matrix
$\mf N_{\varepsilon}$ sharing the same shape with $\mf P$. Therefore, the original sparse core tensor of TAP $\bc{S}$ in (\ref{crit}) becomes $\text{Tensorize} (\text{SparseMax}(\left[\mf N_{\varepsilon}\mf{W}_{Q} \right] \left[ \mf N_{\varepsilon}\mf{W}_{K} \right]^{\T} \oslash \mf{M})$. Similarly, the ablated sparse core tensor of MHTAP is $\text{Tensorize} (\text{SparseMax}(\text{Concat}([\mf{N}_{\varepsilon}\mf{W}_{Q}^{1}][\mf{N}_{\varepsilon}\mf{W}_{K}^{1} ]^{\T}\oslash \mf{M}_1, \cdots, [\mf{N}_{\varepsilon}\mf{W}_{Q}^{h}][\mf{N}_{\varepsilon}\mf{W}_{K}^{h} ]^{\T} \oslash \mf{M}_h)  ))$.
\textit{In this way, the information prior  from partial observations is discarded.}

Table \ref{Tab:ab1} shows the average results of TAP/MHTAP compared to TAP/MHTAP without observation information (OI). The performances  deteriorate significantly in the absence of  the observation information, indicating that the self-supervised prior extracted from the limited observations is useful in  the proposed methods in the 3D physical field reconstruction task.

\begin{table}[t]
	\centering
	\caption{Ablation study on reconstructing radio map 1 with SLNRE metric. All the methods are enhanced by TV regularization.}	
	\renewcommand{\arraystretch}{1.5}
	\begin{tabularx}{0.9\linewidth}{l c c c c }
		\toprule
		Methods (+TV)& $\rho=5\%$& $\rho=10\%$ & $\rho$=$15\%$ &  $\rho=20\%$      \\	
		\midrule
		TAP& \textbf{0.189} &\textbf{0.100} &\textbf{0.073} &\textbf{0.067}\\
		TAP wo OI & 13.75 & 0.911 &0.574 &0.632\\
		\midrule
		MHTAP& \textbf{0.180} &\textbf{0.091} &\textbf{0.071} &\textbf{0.065}\\
		MHTAP wo OI & 0.385 &0.263 & 0.284 &0.261\\
		
		\bottomrule
		\label{Tab:ab1}
	\end{tabularx}
  \vspace{-5mm}
\end{table}

\subsubsection{Impact of SparseMax Function}
\begin{figure}[t]
	\centering
	\begin{subfigure}{0.2\textwidth}
		\includegraphics[width=1\textwidth]{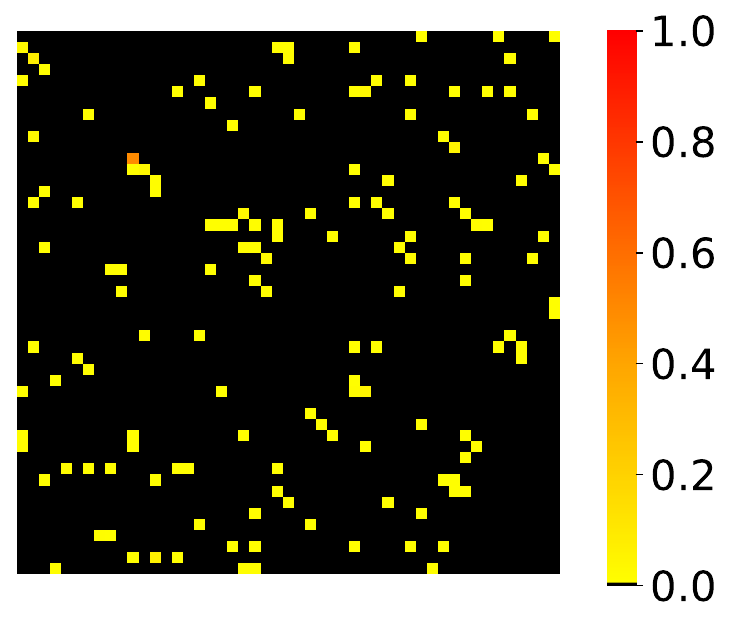}
		\caption{Attention map (SoftMax).}
		\label{fig:sub1}
	\end{subfigure}
	\hspace{0.8cm}
	\begin{subfigure}{0.2\textwidth}
		\includegraphics[width=\textwidth]{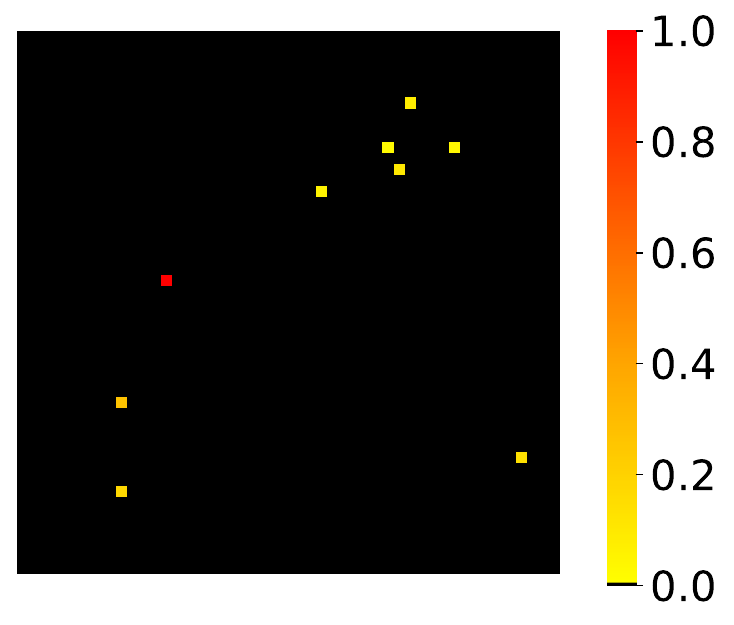}
		\caption{Attention map (SparseMax).}
		\label{fig:sub2}
	\end{subfigure}
	\caption{Partial visulization of attention maps learned by TAP under $\rho = 20\%$ using SoftMax  and SparseMax activation function respectively.}
	\label{fig:am}
\end{figure}
\begin{table}[t]
	\caption{Average reconstruction errors (RMSEs) of TAP and MHTAP using  SparseMax activation or SoftMax activation  versus various $\rho$ values.}
	\renewcommand{\arraystretch}{1.5}
	\begin{tabularx}{\linewidth}{l c c c c }
		\toprule
		Methods& $\rho=5\%$ &  $\rho=10\%$ &$\rho=20\%$ &  $\rho$=$30\%$      \\	
		\midrule
		TAP (SparseMax) & \textbf{1.085} &\textbf{0.564}& \textbf{0.351}&\textbf{0.236}   \\
		TAP (SoftMax)& 1.391& 0.745 &0.363& 0.246  \\
		\midrule
		MHTAP (SparseMax)& \textbf{1.349}&\textbf{0.543}&\textbf{0.319 }  &\textbf{0.215} \\
		MHTAP (SoftMax)&1.626 & 0.714&0.347 &0.231 \\
		\bottomrule
		\label{tab_abla}
	\end{tabularx}
  \vspace{-5mm}
\end{table}
We introduce the SparseMax function into the TAP/MHTAP model to boost the sparsity of the attention map. In this subsection, we compare its performance with the commonly used SoftMax function to demonstrate its effectiveness. The following experiments are conducted on SSF data.
Tab.~\ref{tab_abla} shows the average reconstruction results of TAP/MHTAP using SoftMax  function or using SparseMax function under various $\rho$ values. It indicates  that  the utilization of SparseMax activation enhances performance, particularly when $\rho$ is small. For enhanced visual clarity, we depict $50\times50$ sub-matrices extracted from  attention maps learned by TAP using SoftMax  and SparseMax activation in Fig.~\ref{fig:am}. The attention map learned using the SoftMax function typically consists of many uniformly small scores, whereas the attention map learned using the SparseMax function yields fewer but larger scores. These fewer but larger scores effectively capture the main features of the signals and alleviate overfitting in the reconstruction task.

\section{Conclusion and Future Directions}
\label{sec:6}
In this paper, we introduced FieldFormers based on TAP and MHTAP models for self-supervised reconstruction of 3D physical fields from limited observations. By bridging the Tucker model with the attention scheme, the proposed approach effectively captured both short- and long-range correlations within the limited observations. This learned information was then utilized to adaptively adjust the complexity of the tensor Tucker model, specifically the sparsity of the core tensor. As a result, our method demonstrated promising generalization performance across various types of 3D physical fields. Theoretical analysis was provided to characterize the recovery of ground-truth 3D physical fields using the proposed methods. Additionally, the analysis shed light on the sources of reconstruction errors. Furthermore, we conducted numerical experiments on diverse datasets, including ocean sound speed fields and radio maps, to validate the superiority of the proposed methods in 3D physical field reconstruction compared to SOTA baselines. 
{ While the proposed method has demonstrated strong performance in reconstructing 3D physical fields from limited observations, several directions remain open for future research. In particular, reducing the computational complexity of the tensor attention mechanism would enhance scalability. Also, developing a generative pre-trained version of the proposed approach, which is robust to distribution shifts, could further enhance the reconstruction performance, even under extremely limited observations.}  
\bibliographystyle{IEEEtran}
\bibliography{ref} 

\begin{thebibliography}{10}
\providecommand{\url}[1]{#1}
\csname url@samestyle\endcsname
\providecommand{\newblock}{\relax}
\providecommand{\bibinfo}[2]{#2}
\providecommand{\BIBentrySTDinterwordspacing}{\spaceskip=0pt\relax}
\providecommand{\BIBentryALTinterwordstretchfactor}{4}
\providecommand{\BIBentryALTinterwordspacing}{\spaceskip=\fontdimen2\font plus
\BIBentryALTinterwordstretchfactor\fontdimen3\font minus
  \fontdimen4\font\relax}
\providecommand{\BIBforeignlanguage}[2]{{%
\expandafter\ifx\csname l@#1\endcsname\relax
\typeout{** WARNING: IEEEtran.bst: No hyphenation pattern has been}%
\typeout{** loaded for the language `#1'. Using the pattern for}%
\typeout{** the default language instead.}%
\else
\language=\csname l@#1\endcsname
\fi
#2}}
\providecommand{\BIBdecl}{\relax}
\BIBdecl

\bibitem{Radio}
S.~Shrestha, X.~Fu, and M.~Hong, ``Deep spectrum cartography: Completing radio
  map tensors using learned neural models,'' \emph{IEEE Transactions on Signal
  Processing}, vol.~70, pp. 1170--1184, 2022.

\bibitem{radio_spm}
D.~Romero and S.-J. Kim, ``Radio map estimation: A data-driven approach to
  spectrum cartography,'' \emph{IEEE Signal Processing Magazine}, vol.~39,
  no.~6, pp. 53--72, 2022.

\bibitem{spline}
J.~A. Bazerque, G.~Mateos, and G.~B. Giannakis, ``Group-lasso on splines for
  spectrum cartography,'' \emph{IEEE Transactions on Signal Processing},
  vol.~59, no.~10, pp. 4648--4663, 2011.

\bibitem{hooi}
\BIBentryALTinterwordspacing
L.~Cheng, X.~Ji, H.~Zhao, J.~Li, and W.~Xu, ``{Tensor-based basis function
  learning for three-dimensional sound speed fields},'' \emph{The Journal of
  the Acoustical Society of America}, vol. 151, no.~1, pp. 269--285, 01 2022.
  [Online]. Available: \url{https://doi.org/10.1121/10.0009280}
\BIBentrySTDinterwordspacing

\bibitem{TVT}
Y.~Yue, H.~Zheng, Z.~Shi, and G.~Liao, ``Two-stage reconstruction for co-array
  2d doa estimation of mixed circular and noncircular signals,'' \emph{IEEE
  Transactions on Vehicular Technology}, pp. 1--15, 2025.

\bibitem{TAES}
Y.~Yue, Z.~Zhang, and Z.~Shi, ``Generalized widely linear robust adaptive
  beamforming: A sparse reconstruction perspective,'' \emph{IEEE Transactions
  on Aerospace and Electronic Systems}, vol.~60, no.~5, pp. 5663--5673, 2024.

\bibitem{sound}
R.~Dashen, W.~H. Munk, and K.~M. Watson, \emph{Sound transmission through a
  fluctuating ocean}.\hskip 1em plus 0.5em minus 0.4em\relax Cambridge
  University Press, 2010.

\bibitem{luo}
Z.-Q. Luo, X.~Zheng, D.~López-Pérez, Q.~Yan, X.~Chen, N.~Wang, Q.~Shi, T.-H.
  Chang, and A.~Garcia-Rodriguez, ``Srcon: A data-driven network performance
  simulator for real-world wireless networks,'' \emph{IEEE Communications
  Magazine}, vol.~61, no.~6, pp. 96--102, 2023.

\bibitem{natureoat}
D.~Behringer, T.~Birdsall, M.~Brown, B.~Cornuelle, R.~Heinmiller, R.~Knox,
  K.~Metzger, W.~Munk, J.~Spiesberger, R.~Spindel \emph{et~al.}, ``A
  demonstration of ocean acoustic tomography,'' \emph{Nature}, vol. 299, no.
  5879, pp. 121--125, 1982.

\bibitem{TNN}
S.~Li, L.~Cheng, T.~Zhang, H.~Zhao, and J.~Li, ``Striking the right balance:
  Three-dimensional ocean sound speed field reconstruction using tensor neural
  networks,'' \emph{The Journal of the Acoustical Society of America}, vol.
  154, no.~2, pp. 1106--1123, 2023.

\bibitem{BTD}
G.~Zhang, X.~Fu, J.~Wang, X.-L. Zhao, and M.~Hong, ``Spectrum cartography via
  coupled block-term tensor decomposition,'' \emph{IEEE Transactions on Signal
  Processing}, vol.~68, pp. 3660--3675, 2020.

\bibitem{kri}
W.~C. Van~Beers and J.~P. Kleijnen, ``Kriging interpolation in simulation: a
  survey,'' in \emph{Proceedings of the 2004 Winter Simulation Conference,
  2004.}, vol.~1.\hskip 1em plus 0.5em minus 0.4em\relax IEEE, 2004.

\bibitem{kri2}
G.~Boccolini, G.~Hernandez-Penaloza, and B.~Beferull-Lozano, ``Wireless sensor
  network for spectrum cartography based on kriging interpolation,'' in
  \emph{2012 IEEE 23rd International Symposium on Personal, Indoor and Mobile
  Radio Communications-(PIMRC)}.\hskip 1em plus 0.5em minus 0.4em\relax IEEE,
  2012, pp. 1565--1570.

\bibitem{DL_rm}
S.-J. Kim and G.~B. Giannakis, ``Cognitive radio spectrum prediction using
  dictionary learning,'' in \emph{2013 IEEE Global Communications Conference
  (GLOBECOM)}.\hskip 1em plus 0.5em minus 0.4em\relax IEEE, 2013, pp.
  3206--3211.

\bibitem{AE_rm}
Y.~Teganya and D.~Romero, ``Deep completion autoencoders for radio map
  estimation,'' \emph{IEEE Transactions on Wireless Communications}, vol.~21,
  no.~3, pp. 1710--1724, 2021.

\bibitem{sensor}
X.~Han, L.~Xue, F.~Shao, and Y.~Xu, ``A power spectrum maps estimation
  algorithm based on generative adversarial networks for underlay cognitive
  radio networks,'' \emph{Sensors}, vol.~20, no.~1, p. 311, 2020.

\bibitem{spline1}
J.~Li and A.~D. Heap, ``Spatial interpolation methods applied in the
  environmental sciences: A review,'' \emph{Environmental Modelling \&
  Software}, vol.~53, pp. 173--189, 2014.

\bibitem{TV}
L.~I. Rudin, S.~Osher, and E.~Fatemi, ``Nonlinear total variation based noise
  removal algorithms,'' \emph{Physica D: nonlinear phenomena}, vol.~60, no.
  1-4, pp. 259--268, 1992.

\bibitem{LRTC}
Z.~Zhang, G.~Ely, S.~Aeron, N.~Hao, and M.~Kilmer, ``Novel methods for
  multilinear data completion and de-noising based on tensor-svd,'' in
  \emph{2014 IEEE Conference on Computer Vision and Pattern Recognition}, 2014,
  pp. 3842--3849.

\bibitem{BGCM}
S.~Li, L.~Cheng, T.~Zhang, H.~Zhao, and J.~Li, ``Graph-guided bayesian matrix
  completion for ocean sound speed field reconstruction,'' \emph{The Journal of
  the Acoustical Society of America}, vol. 153, no.~1, pp. 689--710, 2023.

\bibitem{Radiomap}
S.~Bi, J.~Lyu, Z.~Ding, and R.~Zhang, ``Engineering radio maps for wireless
  resource management,'' \emph{IEEE Wireless Communications}, vol.~26, no.~2,
  pp. 133--141, 2019.

\bibitem{K-svd}
\BIBentryALTinterwordspacing
M.~Bianco and P.~Gerstoft, ``{Dictionary learning of sound speed profiles},''
  \emph{The Journal of the Acoustical Society of America}, vol. 141, no.~3, pp.
  1749--1758, 03 2017. [Online]. Available:
  \url{https://doi.org/10.1121/1.4977926}
\BIBentrySTDinterwordspacing

\bibitem{TDL}
\BIBentryALTinterwordspacing
P.~Chen, L.~Cheng, T.~Zhang, H.~Zhao, and J.~Li, ``{Tensor dictionary learning
  for representing three-dimensional sound speed fields},'' \emph{The Journal
  of the Acoustical Society of America}, vol. 152, no.~5, pp. 2601--2616, 11
  2022. [Online]. Available: \url{https://doi.org/10.1121/10.0015056}
\BIBentrySTDinterwordspacing

\bibitem{ICC}
Y.~Teganya and D.~Romero, ``Data-driven spectrum cartography via deep
  completion autoencoders,'' in \emph{ICC 2020-2020 IEEE International
  Conference on Communications (ICC)}.\hskip 1em plus 0.5em minus 0.4em\relax
  IEEE, 2020, pp. 1--7.

\bibitem{quant}
S.~Timilsina, S.~Shrestha, and X.~Fu, ``Quantized radio map estimation using
  tensor and deep generative models,'' \emph{IEEE Transactions on Signal
  Processing}, 2023.

\bibitem{unn}
A.~Qayyum, I.~Ilahi, F.~Shamshad, F.~Boussaid, M.~Bennamoun, and J.~Qadir,
  ``Untrained neural network priors for inverse imaging problems: A survey,''
  \emph{IEEE Transactions on Pattern Analysis and Machine Intelligence},
  vol.~45, no.~5, pp. 6511--6536, 2023.

\bibitem{inductivebias}
J.~Baxter, ``A model of inductive bias learning,'' \emph{Journal of artificial
  intelligence research}, vol.~12, pp. 149--198, 2000.

\bibitem{DIP}
D.~Ulyanov, A.~Vedaldi, and V.~Lempitsky, ``Deep image prior,'' in
  \emph{Proceedings of the IEEE conference on computer vision and pattern
  recognition}, 2018, pp. 9446--9454.

\bibitem{Tennorm}
O.~Semerci, N.~Hao, M.~E. Kilmer, and E.~L. Miller, ``Tensor-based formulation
  and nuclear norm regularization for multienergy computed tomography,''
  \emph{IEEE Transactions on Image Processing}, vol.~23, no.~4, pp. 1678--1693,
  2014.

\bibitem{dp}
Y.~Zheng, J.~Li, Y.~Li, J.~Guo, X.~Wu, and J.~Chanussot, ``Hyperspectral
  pansharpening using deep prior and dual attention residual network,''
  \emph{IEEE transactions on geoscience and remote sensing}, vol.~58, no.~11,
  pp. 8059--8076, 2020.

\bibitem{ml3}
S.~Theodoridis, \emph{Machine Learning: From the Classics to Deep Networks,
  Transformers, and Diffusion Models}.\hskip 1em plus 0.5em minus 0.4em\relax
  Academic Press, 2025.

\bibitem{bengio2021deep}
Y.~Bengio, Y.~Lecun, and G.~Hinton, ``Deep learning for ai,''
  \emph{Communications of the ACM}, vol.~64, no.~7, pp. 58--65, 2021.

\bibitem{SSL}
X.~Liu, F.~Zhang, Z.~Hou, L.~Mian, Z.~Wang, J.~Zhang, and J.~Tang,
  ``Self-supervised learning: Generative or contrastive,'' \emph{IEEE
  transactions on knowledge and data engineering}, vol.~35, no.~1, pp.
  857--876, 2021.

\bibitem{DAP}
W.~He, T.~Uezato, and N.~Yokoya, ``Interpretable deep attention prior for image
  restoration and enhancement,'' \emph{IEEE Transactions on Computational
  Imaging}, vol.~9, pp. 185--196, 2023.

\bibitem{luo2022self}
Y.-S. Luo, X.-L. Zhao, T.-X. Jiang, Y.~Chang, M.~K. Ng, and C.~Li,
  ``Self-supervised nonlinear transform-based tensor nuclear norm for
  multi-dimensional image recovery,'' \emph{IEEE Transactions on Image
  Processing}, vol.~31, pp. 3793--3808, 2022.

\bibitem{lieven}
L.~De~Lathauwer, B.~De~Moor, and J.~Vandewalle, ``A singular value
  decomposition for higher-order tensors,'' in \emph{Second ATHOS workshop,
  Sophia-Antipolis, France}, 1993.

\bibitem{Tucker-als}
N.~D. Sidiropoulos, L.~De~Lathauwer, X.~Fu, K.~Huang, E.~E. Papalexakis, and
  C.~Faloutsos, ``Tensor decomposition for signal processing and machine
  learning,'' \emph{IEEE Transactions on Signal Processing}, vol.~65, no.~13,
  pp. 3551--3582, 2017.

\bibitem{attention}
A.~Vaswani, N.~Shazeer, N.~Parmar, J.~Uszkoreit, L.~Jones, A.~N. Gomez,
  {\L}.~Kaiser, and I.~Polosukhin, ``Attention is all you need,''
  \emph{Advances in neural information processing systems}, vol.~30, 2017.

\bibitem{bert}
J.~Devlin, M.-W. Chang, K.~Lee, and K.~Toutanova, ``Bert: Pre-training of deep
  bidirectional transformers for language understanding,'' \emph{arXiv preprint
  arXiv:1810.04805}, 2018.

\bibitem{maecv}
K.~He, X.~Chen, S.~Xie, Y.~Li, P.~Doll{\'a}r, and R.~Girshick, ``Masked
  autoencoders are scalable vision learners,'' in \emph{Proceedings of the
  IEEE/CVF conference on computer vision and pattern recognition}, 2022, pp.
  16\,000--16\,009.

\bibitem{tong2023bayesian}
X.~Tong, L.~Cheng, and Y.-C. Wu, ``Bayesian tensor tucker completion with a
  flexible core,'' \emph{IEEE Transactions on Signal Processing}, 2023.

\bibitem{TDL2}
F.~Roemer, G.~Del~Galdo, and M.~Haardt, ``Tensor-based algorithms for learning
  multidimensional separable dictionaries,'' in \emph{2014 IEEE international
  conference on acoustics, speech and signal processing (ICASSP)}.\hskip 1em
  plus 0.5em minus 0.4em\relax IEEE, 2014, pp. 3963--3967.

\bibitem{SparseMax}
A.~Martins and R.~Astudillo, ``From softmax to sparsemax: A sparse model of
  attention and multi-label classification,'' in \emph{International conference
  on machine learning}.\hskip 1em plus 0.5em minus 0.4em\relax PMLR, 2016, pp.
  1614--1623.

\bibitem{AF}
\BIBentryALTinterwordspacing
S.~R. Dubey, S.~K. Singh, and B.~B. Chaudhuri, ``Activation functions in deep
  learning: A comprehensive survey and benchmark,'' \emph{Neurocomputing}, vol.
  503, pp. 92--108, 2022. [Online]. Available:
  \url{https://www.sciencedirect.com/science/article/pii/S0925231222008426}
\BIBentrySTDinterwordspacing

\bibitem{kaiming}
K.~He, X.~Zhang, S.~Ren, and J.~Sun, ``Delving deep into rectifiers: Surpassing
  human-level performance on imagenet classification,'' in \emph{2015 IEEE
  International Conference on Computer Vision (ICCV)}, 2015, pp. 1026--1034.

\bibitem{adam}
\BIBentryALTinterwordspacing
A.~Barakat and P.~Bianchi, ``Convergence and dynamical behavior of the adam
  algorithm for nonconvex stochastic optimization,'' \emph{SIAM Journal on
  Optimization}, vol.~31, no.~1, pp. 244--274, 2021. [Online]. Available:
  \url{https://doi.org/10.1137/19M1263443}
\BIBentrySTDinterwordspacing

\bibitem{cn}
Y.~C. Eldar and G.~Kutyniok, \emph{Compressed sensing: theory and
  applications}.\hskip 1em plus 0.5em minus 0.4em\relax Cambridge university
  press, 2012.

\bibitem{ltr}
H.~Rauhut, R.~Schneider, and {\v{Z}}.~Stojanac, ``Low rank tensor recovery via
  iterative hard thresholding,'' \emph{Linear Algebra and its Applications},
  vol. 523, pp. 220--262, 2017.

\bibitem{coveringnumber}
D.-X. Zhou, ``The covering number in learning theory,'' \emph{Journal of
  Complexity}, vol.~18, no.~3, pp. 739--767, 2002.

\bibitem{WC}
A.~Goldsmith, \emph{Wireless communications}.\hskip 1em plus 0.5em minus
  0.4em\relax Cambridge university press, 2005.

\bibitem{flash}
A.~Katharopoulos, A.~Vyas, N.~Pappas, and F.~Fleuret, ``Transformers are rnns:
  Fast autoregressive transformers with linear attention,'' in
  \emph{International conference on machine learning}.\hskip 1em plus 0.5em
  minus 0.4em\relax PMLR, 2020, pp. 5156--5165.

\end{thebibliography}
\qquad
\newpage
\begin{appendix}
	\subsection{Pseudo code of tensorization}
	\label{pseudo code}
	\lstset{ 
		basicstyle=\footnotesize, 
		keywordstyle=\bfseries, 
		language=Python 
	}
\begin{lstlisting}
def tensorize_to_core(s, flag):
if flag == 'TAP': #s:sparse attention map(N*N)
s = s.view(J1, J2, J3, J1, J2, J3)
s = s.permute(0,3,1,4,2,5)
core_tensor = s.reshape(N1, N2, N3) 
elif flag == 'MHTAP': #s:sparse attention map(hN*N)
s = s.view(h1, h2, h3, J1, J2, J3, J1, J2, J3)
s = s.permute(0,3,6,1,4,7,2,5,8)
core_tensor = s.reshape(h1*N1, h2*N2, h3*N3) 
return core_tensor
\end{lstlisting}
	\subsection{3D FieldFormer based on MHTAP.}
	The algorithm is summarized in \textbf{Algorithm \ref{A2}}.
	\label{appendix a}
	\begin{algorithm}[ht]
		\caption{3D FieldFormer based on MHTAP.} 
		\label{A2}
		{\bf Input:} 
		Observations $\bc{Y} \in \mb{R}^{I_1\times I_2\times I_3}$, indicating tensor $\bc{O} \in \mb{R}^{I_1\times I_2\times I_3}$, window size in three mode $(K_1, K_2, K_3)$, stride size in three mode $(S_1, S_2, S_3)$; \\
		{\bf Initialization:} Initialize the query and key matrices $\{\mf{W}_{Q}^i, \mf{W}_{K}^i\}_{i=1}^h$ as well as the value matrices $\mf{V}_1 \in \mb{R}^{I_1\times N_1}, \mf{V}_2 \in \mb{R}^{I_2\times N_2}, \mf{V}_3 \in \mb{R}^{I_3\times N_3}$. 
		\begin{algorithmic}[1]
			\State Extract cubes from observations $\bc{Y}$ to get $\mf{P}$.
			\While{not converge} 
			\State Compute the query-key pairs through  $	\{\mf{Q}_i=\mf{P}\mf{W}_Q^i $, $\mf{K}_i=\mf{P}\mf{W}_K^i\}_{i=1}^h $.
			\State Compute the output of sparse attention module through Eq.~\eqref{eq:sa} and obtain reconstruction  with $\varsigma(\text{MHSTA}(\{\mf{Q}_i\}_{i=1}^h, \{\mf{K}_i\}_{i=1}^h, \mf{V}_1, \mf{V}_2, \mf{V}_3))$.
			\State Compute the  loss $\|\bc{Y} -\bc{O}\ast\bc{X} \|_{\F}^{2}$
			\State Update $\{\mf{W}_{Q}^i, \mf{W}_{K}^i\}_{i=1}^h, \mf{V}_1, \mf{V}_2, \mf{V}_3$ according to the loss using  the Adam optimizer.
			\EndWhile
			\State \textbf{end while}
		\end{algorithmic}
		{\bf Output:} 
		The reconstructed 3D physical field $\bc{X}$.
	\end{algorithm}

	\subsection{{ Implementation details for the SparseMax function}}	
	\label{app:sparsemax}
	{The algorithm is summarized in \textbf{Algorithm \ref{A3}}.}
	\begin{algorithm}[h]
		\caption{{ SparseMax Function.}} 
		\label{A3}
		{{\bf Input:} Vector $\mf{z}=\left[z_{(1)},z_{(2)},\cdots,z_{(N)}\right]\in \mb{R}^{N}$.}\\
		\begin{algorithmic}[1]
			\State  {Sort the elements of $\mathbf{z}$ in descending order:
				\[
				z_{1} \geq z_{2} \geq \dots \geq z_{N}.
				\]}
			\State {Determine the support number $n(\mf{z})$ satisfying:
				\[
				n(\mf{z}) = \max \left\{ n \in \{1, \dots, N\} \mid 1 + n z_{n} > \sum_{j=1}^{n} z_{j} \right\}.
				\]}
			\State {Compute the threshold $\tau(\mf{z})$:
				\[
				\tau(\mf{z})=\frac{\sum_{j\le n(\mf z)} z_j-1 }{n(\mf z)}.
				\]}
			\State  {Compute the final normalized output:
				\[
				p_{(i)} = \max(z_{(i)} - \tau, 0), \quad \forall i \in \{1, \dots, N\}.
				\]}
		\end{algorithmic}
		{{\bf Output:} Normalized output $\mathbf{p} =\left[p_{(1)},p_{(2)},\cdots,p_{(N)}\right]\in \mb{R}^{N}$.} 
	\end{algorithm}

	\subsection{Proof of Lemma 1}
	\label{proof:lemma1}
	Without loss of generality, assume  all entries in $\mf{Q}\mf{K}^{\T}\oslash \mf{M}$ are independent random variables with mean 0 and variance 1. SparseMax applies row-wise sparse normalization and we treat all rows equivalently, so we take a specific row of $\mf{Q}\mf{K}^{\T}\oslash \mf{M}$, denoted by $\mf{z}=\left[z_{(1)},z_{(2)},\cdots,z_{(N)}\right]\in \mb{R}^{N}$ for example. Firstly, SparseMax sorts $\mf{z}$ as $z_1 \le z_2 \le \cdots \le z_N$. Secondly, it compute support number $n(\mf z)=\max \{n \in [N] | 1+nz_n > \sum_{j\le n} z_j  \} $. Thirdly, it computes normalization coefficient $\tau(\mf{z})=\frac{\sum_{j\le n(\mf z)} z_j-1 }{n(\mf z)}$ with $n(\mf z)$ leading scores. Finally, $n(\mf z)$ leading scores are normalized with summation to 1 while others are set to zeroes through threshold function $\left[z_i-\tau(\mf{z}))\right]_{+}$.  It is worthy to note that given a specific number $k$, if $1+kz_k \le \sum_{j\le k} z_j$, then  $1+(k+1)z_{k+1} \le 1+kz_k+z_{k+1} \le \sum_{j\le k} z_j +z_{k+1}=\sum_{j\le k+1} z_j$ holds, which 
	means that if $n=k$ does not satisfy the inequality ($1+nz_n > \sum_{j\le n} z_j$), then any $n>k$ does not satisfy the inequality as well. Therefore,  if $n=k-1$  satisfy the inequality but $n=k$ do not, then $n(\mf{z})=k-1$ holds.
	
	Before computing $\mb{E}(n(\mf z))$, we need to acquire the probability distribution of discrete random variable $n(\mf z)$. Given the above property, we have:
	
	\begin{equation}
		\begin{split}
			&P(n(\mf z)=1)= \noindent\\
			&P(1+z_1 >z_1)P(1+2z_2\le z_1+z_2 | z_2<z_1)=\noindent\\
			&1\times \frac{P(1+z_2 \le z_1, z_2<z_1)}{P(z_2<z_1)}=\frac{P(z_1-z_2 \ge 1)}{P(z_2-z_1<0)}\noindent\\ 
			&= \frac{P( \frac{z_1-z_2}{\sqrt{2}}  \ge \frac{1}{\sqrt{2}})}{P(\frac{z_2-z_1}{\sqrt{2}}<0)}= \frac{1-\Phi(\frac{1}{\sqrt{2}})}{\Phi(0)},
		\end{split}
	\end{equation}
	where $\Phi\left(\cdot \right)$ is the cumulative distribution function (CDF) of Gaussian distribution with mean 0 and variance 1:
	\begin{equation}
		\Phi\left(X \right)= \int_{-\infty  }^{X} \frac{1}{\sqrt{2\pi}}e^{-\frac{x^2}{2}} dx.		
	\end{equation} 
	Similarly, we can get $P(n(\mf z)=2)$:
	\begin{equation}
		\begin{split}
			&P(n(\mf z)=2)= \noindent\\
			&P(1+z_1 >z_1)P(1+2z_2\ge z_1+z_2 | z_2<z_1)\noindent\\
			&P(1+3z_3\le z_1+z_2+z_3 | z_3<z_2)= \noindent\\
			&\frac{P(1+2z_2\ge z_1+z_2, z_2<z_1)}{P(z_2<z_1)}\times \noindent\\
			& \frac{P(1+3z_3\le z_1+z_2+z_3, z_3<z_2)}{P(z_3<z_2)}= \noindent\\
			& \frac{P(0<z_1-z_2<1)}{P(z_2<z_1)}\times \frac{P(z_1+z_2-2z_3\ge 1)}{P(z_3<z_2)}= \noindent\\
			&\frac{\Phi(\frac{1}{\sqrt{2}})-\Phi(0) }{\Phi(0)}\times \frac{1-\Phi(\frac{1}{2}) }{\Phi(0)}.
		\end{split}
	\end{equation}
	To conclude, for $N>3$, we have:
	\begin{equation}
		\begin{split}
			\label{eq:prob}
			&P(n(\mf z)=1)=2-2\Phi(\frac{1}{\sqrt{2}}),\noindent \\ 
			&P(n(\mf z)=n)=(2-2\Phi(\frac{1}{\sqrt{2n}  })) \times \prod_{i=2}^{n}(2\Phi(\frac{1}{2i-2})-1),\noindent\\
			&~~~~~~~~~~~~~~~~~~~~~~~~~~~~~~~~~~~~~~~~~~~~n = 2,\cdots, N-1,\noindent\\
			&P(n(\mf z)=N)=\prod_{i=2}^{N}(2\Phi(\frac{1}{\sqrt{2i-2} })-1 ).
		\end{split}
	\end{equation}
	
	We can infer from Eq.~\eqref{eq:prob} that $P(n(\mf{z}))$ is a constant independent of $N$ and that most of the probability density falls into small values. And  the expectation of $n(\mf z)$ can be computed through $\mb{E}(n(\mf z))=\sum_{i}^{N}iP(n(\mf z)=i)$. Therefore, the expectation  of the number of the non-zero elements in attention map (i.e.,  $\|\bc{S}\|_{0}$) equals $N\mb{E}(n(\mf z))$.

	\subsection{Proof of Lemma 2}
	\label{proof:lemma2}
	\textit{Lemma 4}\cite{cn} \textit{:} Let $\mathcal{Z}$ be a subset of a vector space of real dimension $R$ with unit norm ($\left\|\mathcal{Z}\right\|_{\F} = 1$) and let $0<\varepsilon<1$. The covering number of $\mathcal{Z}$ scaled by $\varepsilon$ is upper bounded by:
	\begin{equation}
		\mathsf N(\mathcal{Z},\varepsilon) \le \left(\frac{3}{\varepsilon}\right)^R.
	\end{equation}
	\hfill  $\blacksquare$

	Consider the set $\mathcal{S}=\{\bc{S}\in \mb{R}^{N_1 \times N_2 \times N_3}, \left\|\bc{S}\right\|_{\F} \le \alpha \}$ containing the core tensor and $\mathcal{V}_1 = \{\mf{V}_1\in \mb{R}^{I_1 \times N_1}, \left\|\mf{V}_1\right\|_{\F} \le \beta\}, \mathcal{V}_2 = \{\mf{V}_2\in \mb{R}^{I_2 \times N_2}, \left\|\mf{V}_2\right\|_{\F} \le \beta\}, \mathcal{V}_3 = \{\mf{V}_3\in \mb{R}^{I_3 \times N_3}, \left\|\mf{V}_3\right\|_{\F} \le \beta\} $ containing the factor matrices.
	Since $\mathcal{S}$ is an Euclidean ball that has a radius of $\alpha$ in the $\|\bc{S}\|_{0}$-dimensional space.
	Then, according to \textit{Lemma 4},  the covering number of $\mathcal{S}$ is upper bounded by :
	\begin{equation}
		\mathsf N(\mathcal{S},\varepsilon) \le \left(\frac{3\alpha}{\varepsilon}\right)^{\|\bc{S}\|_{0}}.
	\end{equation}
	The above leads to the following upper bound:
	\begin{equation}
		\mathsf	N(\mathcal{S}, \frac{\varepsilon}{\beta^3+3\alpha\beta^2}) \le \left(\frac{3\alpha(\beta^3+3\alpha\beta^2)}{\varepsilon}\right)^{\|\bc{S}\|_{0}}.
	\end{equation}
	
	Also,  $\mathcal{V}_l$ is an Euclidean ball that has a radius of $\beta$ in the $I_lN_l$-dimensional space.
	Similarly, the covering number of $\mathcal{V}_l$ is upper bounded by :
	\begin{equation}
		\mathsf	N(\mathcal{V}_l,\varepsilon) \le  \left(\frac{3\beta}{\varepsilon}\right)^{I_lN_l},
	\end{equation}
	for $i=1,2,3$. This also holds that:
	\begin{equation}
		\mathsf	N(\mathcal{V}_l,\frac{\varepsilon}{\beta^3+3\alpha\beta^2}) \le  \left(\frac{3\beta(\beta^3+3\alpha\beta^2)}{\varepsilon}\right)^{I_lN_l},
	\end{equation}
	for $i=1,2,3$. Here we define the set $\mathcal{X}_{} = \{\tilde{\bc{X}} =  \tilde{\bc{S}} \times_1 \tilde{\mf{V}}_1 \times_2 \tilde{\mf{V}}_2 \times_3 \tilde{\mf{V}}_3| \tilde{\bc{S}}\in \mathcal{S}, \{\tilde{\mf{V}}_l \in \mathcal{V}_l\}_{i=1}^{3}\}$.
	
	Let $\bar{\mathcal{S}}$ be an $\frac{\varepsilon}{\beta^3+3\alpha\beta^2}$-net of $\mathcal{S}$ and $ \bar{\mathcal{V}}_l$ be an $\frac{\varepsilon}{\beta^3+3\alpha\beta^2}$-net of $\mathcal{V}_l$ for $i=1,2,3$, respectively. And we define the set $\bar{\mathcal{X}} = \{\bar{\bc{X}} =  \bar{\bc{S}} \times_1 \bar{\mf{V}}_1 \times_2 \bar{\mf{V}}_2 \times_3 \bar{\mf{V}}_3| \bar{\bc{S}}\in \bar{\mathcal{S}}, \{\bar{\mf{V}}_l \in  \bar{\mathcal{V}}_l\}_{l=1}^{3}\}$. Hence, the cardinality of $\bar{\mathcal{X}}_{}$ is upper bounded by
	\begin{equation}
		\begin{split}
			&|\bar{\mathcal{X}}_{}| \le \\ &\left(\frac{3\alpha(\beta^3+3\alpha\beta^2)}{\varepsilon}\right)^{\|\bc{S}\|_{0}}\prod_{i=1}^{3} \left(\frac{3\beta(\beta^3+3\alpha\beta^2)}{\varepsilon}\right)^{I_lN_l} \\
			&=    	\left[\frac{3(\beta^{3} +3\alpha\beta^{2})}{\varepsilon}\right]^{\|\bc{S}\|_{0}+\sum_{l=1}^{3}N_lI_l }\alpha^{\|\bc{S}\|_{0}} \beta^{\sum_{l=1}^{3}N_lI_l}	
		\end{split}
	\end{equation}

	Consider  $\tilde{\bc{X}}\in \mathcal{X}_{}$  and  $\bar{\bc{X}}\in \bar{\mathcal{X}}_{}$,  which can be represented as $\tilde{\bc{X}} = \tilde{\bc{S}} \times_1 \tilde{\mf{V}}_1 \times_2 \tilde{\mf{V}}_2 \times_3 \tilde{\mf{V}}_3, \tilde{\bc{S}}\in \mathcal{S}, \{\tilde{\mf{V}}_{l}  \in \mathcal{V}_l\}_{l=1}^3$  and $\bar{\bc{X}} = \bar{\bc{S}} \times_1 \bar{\mf{V}}_1 \times_2 \bar{\mf{V}}_2 \times_3 \bar{\mf{V}}_3, \bar{\bc{S}}\in \bar{\mathcal{S}}, \{\bar{\mf{V}}_{i}  \in  \bar{\mathcal{V}}_l\}_{l=1}^3$ respectively.
	Hence, for any $\tilde{\bc{X}}$, there exist $\bar{\bc{X}}$ such that the following holds:
	\begin{equation}
		\begin{split}
			\|\tilde{\bc{X}}-&\bar{\bc{X}}\|_{\F}=\left\|\tilde{\bc{S}} \times \tilde{\mf{V}}_1 \times \tilde{\mf{V}}_2 \times \tilde{\mf{V}}_3-\bar{\bc{S}} \times \bar{\mf{V}}_1 \times \bar{\mf{V}}_2 \times \bar{\mf{V}}_3 \right\|_{\F}\\
			=&\|\tilde{\bc{S}} \times_1 \tilde{\mf{V}}_1 \times_2 \tilde{\mf{V}}_2 \times_3 \tilde{\mf{V}}_3 \pm  \tilde{\bc{S}} \times_1 \tilde{\mf{V}}_1 \times_2 \tilde{\mf{V}}_2 \times_3 \bar{\mf{V}}_3  \pm \\ &\tilde{\bc{S}} \times_1 \tilde{\mf{V}}_1 \times_2 \bar{\mf{V}}_2 \times_3 \bar{\mf{V}}_3 \pm  \tilde{\bc{S}} \times_1 \bar{\mf{V}}_1 \times_2 \bar{\mf{V}}_2 \times_3 \bar{\mf{V}}_3  \\ & -\bar{\bc{S}} \times_1 \bar{\mf{V}}_1 \times_2 \bar{\mf{V}}_2 \times_3 \bar{\mf{V}}_3\|_{\F}\\
			\le&  \|\tilde{\bc{S}} \times \tilde{\mf{V}}_1 \times \tilde{\mf{V}}_2 \times (\tilde{\mf{V}}_3-\bar{\mf{V}}_3) \|_{\F} + \\
			& \|\tilde{\bc{S}} \times \tilde{\mf{V}}_1 \times (\tilde{\mf{V}}_2-\bar{\mf{V}}_2) \times \bar{\mf{V}}_3 \|_{\F}+\\
			& \|\tilde{\bc{S}} \times (\tilde{\mf{V}}_1-\bar{\mf{V}}_1) \times \bar{\mf{V}}_2 \times \bar{\mf{V}}_3 \|_{\F}+\\
			& \|(\tilde{\bc{S}}-\bar{\bc{S}}) \times \bar{\mf{V}}_1 \times \bar{\mf{V}}_2 \times \bar{\mf{V}}_3 \|_{\F}\\
			\le&3\alpha\beta^2\frac{\varepsilon}{\beta^3+3\alpha\beta^2}+\beta^3\frac{\varepsilon}{\beta^3+3\alpha\beta^2}=\varepsilon.
		\end{split}
	\end{equation}
	Therefore, $\bar{\mathcal{X}}$  is an $\varepsilon$-net of $\mathcal{X}$ with a  covering number is upper bounded by
	\begin{equation}
		\begin{split}    	
			\mathsf N&(\mathcal{X}_{},\varepsilon) \le  \\
			&\left[\frac{3(\beta^{3} +3\alpha\beta^{2})}{\varepsilon}\right]^{\|\bc{S}\|_{0}+\sum_{l=1}^{3}N_lI_l }\alpha^{\|\bc{S}\|_{0}} \beta^{\sum_{l=1}^{3}N_lI_l}	
		\end{split}
	\end{equation}
	Next, consider the $T$-Lipschitz continuous activation function $\varsigma(\cdot)$ such that the following holds for any $\tilde{\bc{X}}_1, \tilde{\bc{X}}_2 \in \mathcal{X}$:
	\begin{equation}
		\|\varsigma(\tilde{\bc{X}}_1)-\varsigma(\tilde{\bc{X}}_2)\|_{\F} \le T\|\tilde{\bc{X}}_1-\tilde{\bc{X}}_2\|_{\F}
	\end{equation}
	Consider the set  $\mathcal{X}_{\text{TAP}} = \{\varsigma(\tilde{\bc{X}}) = \varsigma(\tilde{\bc{S}} \times_1 \tilde{\mf{V}}_1 \times_2 \tilde{\mf{V}}_2 \times_3 \tilde{\mf{V}}_3)| \tilde{\bc{S}}\in \mathcal{S}, \{\tilde{\mf{V}}_l \in \mathcal{V}_l\}_{i=1}^{3}\}$, we find that an $\frac{\varepsilon}{T}$-net of $\mathcal{X}$ is an $\varepsilon$-net of   $\mathcal{X}_{\text{TAP}}$. Hence, we have 
	\begin{equation}
		\begin{split}    	
			\mathsf N&(\mathcal{X}_{\text{TAP}},\varepsilon) \le  \\
			&\left[\frac{3T(\beta^{3} +3\alpha\beta^{2})}{\varepsilon}\right]^{\|\bc{S}\|_{0}+\sum_{l=1}^{3}N_lI_l }\alpha^{\|\bc{S}\|_{0}} \beta^{\sum_{l=1}^{3}N_lI_l}.	
		\end{split}
	\end{equation}
	This completes the proof.

	\subsection{Proof of Lemma 3}
	\label{proof:lemma3}
	\textit{Lemma 5:}\cite{Radiomap} Let $\mathcal{Y}_1, \cdots,\mathcal{Y}_w$ be a set of samples taken without replacement from $\left\{y_1,\cdots,y_n\right\}$ of mean $\mu$. Denote $a = \min_{i}y_i$ and $b = \max_iy_i$, then
	\begin{equation}
		\text{Pr}\left[|\frac{1}{w}\sum_{i=1}^{w}\mathcal{Y}_i - \mu|\ge t \right] \le 2 \exp \big(-\frac{2wt^2}{(1-(w-1)/n) (b-a)^2}\big).
	\end{equation}
	\hfill  $\blacksquare$
	
	Consider a set of variables ${\mathcal{W}(i_1,i_2,i_3) = (\tilde{\bc{Y}}(i_1,i_2,i_3) - \tilde{\bc{X}}(i_1,i_2,i_3))^2, \forall(i_1,i_2,i_3) \in \Omega }$. Then, the sample mean of $\mathcal{W} (i_1,i_2,i_3)$ is the empirical loss $\text{loss}_1(\tilde{\bc{X}})$ and the actual mean is $\text{loss}_2(\tilde{\bc{X}})$. One can see that $\mathcal{W}(i_1,i_2,i_3) = (\tilde{\bc{Y}}(i_1,i_2,i_3) - \tilde{\bc{X}}(i_1,i_2,i_3))^2 \le (|\tilde{\bc{Y}}(i_1,i_2,i_3)|+|\tilde{\bc{X}}(i_1,i_2,i_3)|)^2 \le (\nu + v + \alpha\beta^3)^2 = \xi$ with $\nu = \max_{i_1,i_2,i_3} \left|\bc{X}_{\natural}(i_1,i_2,i_3)  \right| $ and $v = \max_{i_1,i_2,i_3} \left|\bc{N}(i_1,i_2,i_3)  \right| $.
	Using \textit{Lemma 5}, we have
	$\text{Pr}\left[|\text{loss}_1(\tilde{\bc{X}}) - \text{loss}_2(\tilde{\bc{X}})|   \ge t \right] \le 2 \exp \big(-\frac{2|\Omega|t^2}{(1-(|\Omega|-1)/I_1I_2I_3) \xi^2}\big).$ Denote an $\varepsilon$-net of $\mathcal{X}_{\text{TAP}}$ as $\bar{\mathcal{X}}_{\text{TAP}}$. Using the union bound on all $\bar{\bc{X}}\in \bar{\mathcal{X}}_{\text{TAP}}$ yields $\text{Pr}\left[\sup_{\bar{\bc{X}} \in \bar{\mathcal{X}}_{\text{TAP}} } |\text{loss}_1(\bar{\bc{X}}) - \text{loss}_2(\bar{\bc{X}})|   \ge t \right] \le 2 |\bar{\mathcal{X}}_{\text{TAP}}| \exp \big(-\frac{2|\Omega|t^2}{(1-(|\Omega|-1)/I_1I_2I_3) \xi^2}\big).$ Equivalently, with probability at least $1-\delta$, the following holds:
	\begin{equation}
		\begin{split}
			&\sup_{\bar{\bc{X}}\in \bar{\mathcal{X}}_{\text{TAP}} }|\text{loss}_1(\bar{\bc{X}}) - \text{loss}_2(\bar{\bc{X}})|\le \\
			&\sqrt{\frac{\xi^2\log(2|\bar{\mathcal{X}}_{\text{TAP}}|/\delta)}{2}(\frac{1}{|\Omega|}+\frac{1}{|\Omega|I_1I_2I_3} - \frac{1}{I_1I_2I_3})}.
		\end{split}
	\end{equation}
	Let $u(\Omega)  \triangleq \sup_{\bar{\bc{X}}\in \bar{\mathcal{X}}_{\text{TAP}} }|\text{loss}_1(\bar{\bc{X}}) - \text{loss}_2(\bar{\bc{X}})| $. We have $\sup_{\bar{\bc{X}}\in \bar{\mathcal{X}}_{\text{TAP}} }| \sqrt{\text{loss}_1(\bar{\bc{X}})} - \sqrt{\text{loss}_2(\bar{\bc{X}})}| \le \sqrt{u(\Omega)}$.

	Note that
	\begin{equation}
		\begin{split}
			&|\sqrt{\text{loss}_2(\tilde{\bc{X}})} - \sqrt{\text{loss}_2(\bar{\bc{X}})} | =\\
			& \frac{1}{\sqrt{I_1I_2I_3}}|\|\tilde{\bc{Y}}-\tilde{\bc{X}}\|_{\F}-\|\tilde{\bc{Y}}-\bar{\bc{X}}\|_{\F}| \\
			&\le \frac{1}{\sqrt{I_1I_2I_3}}\|\tilde{\bc{Y}}-\tilde{\bc{X}}-\tilde{\bc{Y}}+\bar{\bc{X}}\|_{\F} = \frac{1}{\sqrt{I_1I_2I_3}}\|\bar{\bc{X}}-\tilde{\bc{X}} \|_{\F}\\
			& \le \frac{\varepsilon}{\sqrt{I_1I_2I_3}}.
		\end{split}
	\end{equation}

	Similarly, we have $|\sqrt{\text{loss}_1(\tilde{\bc{X}})} - \sqrt{\text{loss}_1(\bar{\bc{X}})}| \le \frac{\varepsilon}{\sqrt{|\Omega|} }$. Then we can show that the following holds:
	\begin{equation}
		\begin{split}
			&\sup_{\tilde{\bc{X}}\in \mathcal{X}_{\text{TAP}}}|\sqrt{\text{loss}_1(\tilde{\bc X})} - \sqrt{\text{loss}_2(\tilde{\bc X})}| \le \\
			& \frac{\varepsilon}{\sqrt{I_1I_2I_3}} + \sqrt{u(\Omega)} + \frac{\varepsilon}{\sqrt{|\Omega|}}.
		\end{split}
	\end{equation}
	Therefore, the following holds using the definition of $u(\Omega)$ with probability at least $1-\delta$:
	\begin{equation}
		\begin{split}
			\sup_{\tilde{\bc{X}}\in \mathcal{X}_{\text{TAP}}}&|\sqrt{\text{loss}_1(\tilde{\bc X})} - \sqrt{\text{loss}_2(\tilde{\bc X})}| \le \\  &\frac{2\varepsilon}{\sqrt{|\Omega|}}+\big(\frac{\xi^2 w}{2}\log(\frac{2}{\delta}|\bar{\mathcal{X}}_{\text{TAP}}|)\big)^{\frac{1}{4}}.
		\end{split}
	\end{equation}
	This completes the proof.

	\subsection{Proof of Theorem 1}
	\label{proof:theorem1}
	Denote the empirical loss and actual loss associated with an optimal solution $\bc{X}^{\ast}$ as:
	\begin{equation}
		\begin{split}
			&\sqrt{\text{loss}_1(\bc{X}^{\ast})} = \frac{1}{\sqrt{|\Omega|} }  \|\bc{O} \ast (\tilde{\bc{Y}} - \bc{X}^{\ast})\|_{\F},\\
			&\sqrt{\text{loss}_2(\bc{X}^{\ast})} = \frac{1}{\sqrt{I_1I_2I_3} }  \|\tilde{\bc{Y}} - \bc{X}^{\ast}\|_{\F}.
		\end{split}
	\end{equation}
	Then the following chain of inequality holds:
	\begin{equation}
		\begin{split}
			&\frac{1}{\sqrt{I_1I_2I_3}}\|\bc{X}^{\ast}-\bc{X}_{\natural}\|_{\F} =  \frac{1}{\sqrt{I_1I_2I_3}}\|\bc{X}^{\ast}-\tilde{\bc{Y}}+ \bc{N}\|_{\F}\\
			&\le \frac{1}{\sqrt{I_1I_2I_3}}\|\bc{X}^{\ast}-\tilde{\bc{Y}}\|_{\F} + \frac{1}{\sqrt{I_1I_2I_3}}\|\bc{N}\|_{\F}\\
			&\le  \frac{1}{ \sqrt{|\Omega|} }  \|\bc{O} \ast (\tilde{\bc{Y}} - \bc{X}^{\ast})\|_{\F} + \text{Gap}^{\ast}(\Omega) +\frac{1}{\sqrt{I_1I_2I_3}}\|\bc{N}\|_{\F}\\
			&\le \frac{1}{\sqrt{|\Omega |} }  \|\bc{O} \ast (\tilde{\bc{Y}} - \tilde{\bc{X}}^{\ast})\|_{\F} + \text{Gap}^{\ast}(\Omega) +\frac{1}{\sqrt{I_1I_2I_3}}\|\bc{N}\|_{\F}\\
			& \le \frac{\|\bc{O}\ast(\tilde{\bc{X}}^{\ast}-\bc{X}_{\natural})\|_{\F} + \|\bc{O}\ast (\bc{X}_{\natural}-\tilde{\bc{Y}})\|_{\F}} {\sqrt{|\Omega|}}\\
			&+ \text{Gap}^{\ast}(\Omega) +\frac{1}{\sqrt{I_1I_2I_3}}\|\bc{N}\|_{\F}\\	
			& \le \frac{1}{\sqrt{\left | \Omega \right | }}\left\| \tilde{\bc{X}}^{\ast} - \bc{X}_{\natural} \right\|_{\F} +  \frac{1}{\sqrt{\left | \Omega \right | }} \left\|\bc{O} \ast \bc{N} \right\|_{\F}\\
			&+\text{Gap}^{\ast}(\Omega) + \frac{1}{\sqrt{I_1I_2I_3}}\left\| \bc{N}\right\|_{\F}.
		\end{split}
	\end{equation}
	This completes the proof.
	
	\subsection{ Comparisons of  recoverability analysis}
	\label{app:comparison}
	{ 
		The key differences between the proposed recoverability analysis and prior work lie in:
		\begin{itemize}
			\item The use of a nonlinear Tucker model with an activation function.
			\item The presence of a core tensor with a sparsity pattern.
			\item The use of non-orthogonal factor matrices.
		\end{itemize}
		
		These differences primarily influence the generalization error (i.e., $\text{Gap}^{\ast}(\Omega)$)  in the reconstruction error bound.  In [1], the generalization error is given by:
		\begin{equation}
			\text{Gap}^{\ast}(\Omega) \le \frac{2cR}{\sqrt{|\Omega|}}+\big(\frac{\xi^2\omega}{2}\log(\frac{2\mathsf{N}(\mathcal{X}_{\mathbf{g}R,\boldsymbol{\theta}_d},cR)}{\delta})\big),
		\end{equation}
		with the covering number 
		\begin{align}
			\mathsf{N}(\mathcal{X}_{\mathbf{g}R,\boldsymbol{\theta}_d},\varepsilon) \le (\frac{3R(\alpha+\beta)}{\varepsilon})^{R(K+D)} \alpha^{RK}(Pq)^{RD}. 
		\end{align}
		In contrast, the proposed model yields:
		\begin{align}
			&\text{Gap}^{\ast}(\Omega) \le \frac{2\varepsilon}{\sqrt{|\Omega|}}+\big(\frac{\xi^2\omega}{2}\log(\frac{2\mathsf{N}(\mathcal{X}_{\text{{TAP}}},\varepsilon)}{\delta})\big), \\
			&\mathsf{N}(\mathcal{X}_{\text{TAP}},\varepsilon) \le \nonumber \\ &\left[\frac{3T(\beta^3+3\alpha\beta^{2})}{\varepsilon} \right]^{\|\bc{S}\|_{0}+\sum_{l=1}^{3}N_lI_l}\alpha^{\|\bc{S}\|_{0}}\beta^{\sum_{l=1}^{3}N_lI_l}.
		\end{align}
		Comparing (53)-(56), we observe that the key difference in the generalization error arises from the covering numbers $\mathsf{N}(\mathcal{X}_{\mathbf{g}R,\boldsymbol{\theta}_d},\varepsilon)$ and $\mathsf{N}(\mathcal{X}_{\text{TAP}},\varepsilon)$. In the SOTA model [1], the covering number is determined by pre-selected hyperparameters, such as the assumed source number, $R$ and the latent space dimensionality, $D$. Choosing overly large values increases the bound, thereby raising the risk of higher generalization error.   In contrast, the proposed model's generalization error depends on the sparsity of the core tensor, $\|\bc{S}\|_{0}$, which represents the number of non-zero elements in the core tensor. Instead of being pre-fixed, $\|\bc{S}\|_{0}$ is learned in a self-supervised manner during training. This highlights the importance of learning the core tensor sparsity and aligns with the experimental results.

		Additionally, these differences also influence the representation error in the reconstruction error bound:
		\begin{equation}
			\|\tilde{\bc{X}}^{\ast} - \bc{X}_{\natural}\|_{\F}, 
		\end{equation}
		where $\tilde{\bc{X}}^{\ast}$ represents the best reconstruction from the solution set (see Definition 1) and $\bc{X}_{\natural}$ denotes the ground truth field. In [1], the optimal reconstruction is given by:
		\begin{equation}
			\tilde{\bc{X}}^{\ast} = \arg\min_{\tilde{\bc{X}} \in \mathcal{X}_{\mathbf{g}R,\boldsymbol{\theta}_d}} \|\tilde{\bc{X}} - \bc{X}_{\natural}\|_{\F}, 
		\end{equation}
		while in the proposed model:
		\begin{equation}  
			\tilde{\bc{X}}^{\ast} = \arg\min_{\tilde{\bc{X}} \in \mathcal{X}_{\text{TAP}}} \|\tilde{\bc{X}} - \bc{X}_{\natural}\|_{\F}. 
		\end{equation} 
		
		For the model in [1], the representation error is determined by the pretrained deep model with parameters $\boldsymbol{\theta}_d$. If the test data distribution differs significantly from the training data, the prior information embedded in the pretrained model may misleadingly constrain the solution set, resulting in increased representation error. In contrast, the proposed model does not rely on pretrained models, and its high expressiveness -- since the Tucker model is a universal approximator -- enables a lower representation error.

		In summary, by comparing the two primary sources that contribute to the reconstruction error -- the generalization error in (53)-(56) and the representation error in (57)-(59) -- we conclude that the proposed model has the potential to achieve a lower reconstruction error compared to [1], particularly under distribution shifts. This advantage stems from the proposed model's ability to adaptively adjust its complexity by learning the sparsity of the core tensor, unlike the fixed model complexity in \cite{Radio}, which depends on a pretrained deep model.
		
	}

	\subsection{Implementation Details of Ocean Sound Speed Field Reconstruction}
	\label{idssf}
	\begin{enumerate}
		\item[a.]  Tucker-ALS: The core tensor size is set to (5, 5, 5), and the size of the three factor matrices is set to (20, 5)
		\item[b.] LRTC:  The hyperparameters $\alpha$ and $\beta$ are both set to 1.
		\item[c.] TNN: We choose the  TNN  with layer dimensionalities of (5, 5, 5), (10, 10, 10), and (20, 20, 20), which have been shown to be the most competent  in the SSF reconstruction task\cite{TNN}. The learning rate is set to $4e^{-3}$. The number of parameters is 875. 
		\item[d.] TAP: The cube size is set to (4, 4, 4) and the stride is set to (2, 2, 2). Consequently,  $N = (\frac{20-4}{2}+1)^{3} = 729$ cubes can be extracted, and the reshape size ($N_1, N_2, N_3$) is (81, 81, 81). The learning rate is set to $4e^{-3}$. The number of parameters is 12K.
		\item[e.] MHTAP: The cube size is set to (5, 5, 5) and the stride is set to (3, 3, 3). This configuration allows us to extract a total of $N = (\frac{20-5}{3}+1)^{3}=216$ cubes, with $M = 64$. We choose the head number $h$ to be 8 ($h_1=h_2=h_3=2$) and set the reshape size ($h_1N_1, h_2N_2, h_3N_3$) to $(72, 72, 72)$, which approximates the reshape size used in TAP. The learning rate is set to $4e^{-3}$  and the number of parameters is 132K.
	\end{enumerate}
	
	\subsection{Implementation Details of Radio Map Reconstruction}
	\label{idrm}
	\begin{enumerate}
		\item[a.] TNN: We choose the  TNN  with dimensionality of each layer being (10,10,10), (25,25,32), (51,51,64). The learning rate is set to $4e^{-3}$. The number of parameters is 6.5K. 
		
		\item[b.] TAP: The cube size is set to (6,6,8), the stride is set to (5,5,8). Therefore,  $N = (\frac{51-6}{5}+1)^{2}(\frac{64-8}{8}+1)=800$ cubes  can be  extracted and $M = 288$. The  size of core tensor ($N_1, N_2, N_3$) = (100,100,64). The learning rate is set to $4e^{-3}$. The number of parameters is 180K.
		\item[c.] MHTAP: The cube size is set to (9,9,8), the stride is set to (7,7,8). Therefore,  $N = (\frac{51-9}{7}+1)^{2}(\frac{64-8}{8}+1)=392$ cubes  can be  extracted and $M = 288$. We choose the head number $h$ = 4 and we set the  size of core tensor ($h_1N_1, h_2N_2, h_3N_3$) = $(98,98,64)$ approximating that of TAP.  The learning rate is set to $4e^{-3}$.  The number of parameters is 1.5M.
	\end{enumerate}

	\subsection{Enhanced Loss Function with TV Regularization for Untrained Methods}
	\label{app:loss_tv}
	
	To further enhance the  competitiveness of those untrained models, we incorporate the commonly used total variation (TV) regularization\cite{TV} into the loss function to  enhance the performance. The second order TV regularization for 3D physical fields is defined as:
	\begin{equation}
		\begin{split}
			\|\bc{X}\|_{\text{TV}}&= \sum_{k} \sum_{i,j} \sqrt{D_{x}^2 + D_{y}^2 + 2D_{x,y}^2} \\
			D_{x} =& (\bc{X}(i_1+1,i_2,i_3)-2\bc{X}(i_1,i_2,i_3)+\bc{X}(i_1-1,i_2,i_3))^2 \\
			D_{y} =& (\bc{X}(i_1,i_2+1,i_3)-2\bc{X}(i_1,i_2,i_3)+\bc{X}(i_1,i_2-1,i_3))^2 \\
			D_{x,y} =& \frac{1}{4}(\bc{X}(i_1+1,i_2+1,i_3)- \bc{X}(i_1+1,i_2-1,i_3)- \\
			&\bc{X}(i_1-1,i_2+1,i_3)+\bc{X}(i_1-1,i_2-1,i_3))^2
		\end{split}
	\end{equation}
	Then the loss function of those untrained methods becomes:
	\begin{equation}
		\begin{split}
			&\min_{\bc{X}} \|\bc{Y} - \bc{O}\ast \bc{X}  \|_{\F}^{2} + \gamma\|\bc{X}\|_{\text{TV}},\\
		\end{split}
	\end{equation}
	where  $\bc{X}$ is the reconstructed 3D physical field and $\gamma$ is the trade-off parameter.

	\subsection{ Generic guidelines for the choice of hyperparameters}
	\label{app:L}
	{Here, we provide guidelines for selecting dimensions ($R_1, R_2, R_3$),
		($K_1,K_2,K_3$) and ($S_1,S_2,S_3$). 
		
		Our key idea is to first construct an over-complete tensor Tucker model with a Tucker core larger than the physical field to ensure expressiveness (i.e., $R_l > I_l, \forall l$). We then employ an attention mechanism to adaptively learn the sparsity pattern (model complexity). To achieve this, we recommend choosing  $K_l \le \lfloor\frac{I_l}{3} \rfloor, \forall l$, where $K_l$ and $I_l$ denote the  window size and the original field size for mode $l$, respectively. The stride size controls the number of extracted cubes, and we suggest choosing $S_l \le \lfloor\frac{K_l}{2} \rfloor, \forall l$. 
		
		Once $(K_1,K_2,K_3)$ and $(S_1,S_2,S_3)$ are selected, the size of the core tensor, $(R_1,R_2,R_3)$, can be determined by the proposed tensorization steps in Sec. III-B,  ensuring that $R_l > I_l, \forall l$ holds.  Additionally, the empirical results indicate that the corresponding final reconstruction remains insensitive to different hyperparameter choices, provided the specified conditions   (i.e., $R_l > I_l, \forall l$) are met.}

	\subsection{ Impact of the sparsity level and pattern of the observations}
	\label{app:M}
	{ Since our goal is to develop a ``complexity-adaptive'' model for general physical field reconstructions, our derivation of the reconstruction error bound (Theorem 1) imposes minimal restrictions on the sparsity levels ($\rho$) and the  patterns in the original observation set, $\bc{Y}$. Specifically, the sparsity level $\rho$ primarily influences the error bound through the term $| \Omega |$ in Eq. (26) and Eq. (27). A higher $\rho$ (i.e., larger $| \Omega |$) reduces the estimation error bound, as shown in Eq. (27).
		
		Although the sparsity pattern of $\bc{Y}$ is not explicitly included in the recoverability analysis, we conduct experiments with different sampling patterns to evaluate the reconstruction performance under practical scenarios. In these experiments, we use random sampling for SSF and fiber-wise sampling for the radio map. Empirically, the proposed model performs well with both sampling strategies, demonstrating its robustness.
		
		While our derived error bound does not explicitly quantify the impact of the sparsity pattern of $\bc{Y}$, it does influence reconstruction results to some extent. This presents an interesting direction for future work -- to refine the error bound by incorporating sparsity patterns. In particular, adversarial sampling scenarios (e.g., block-missing observations) can degrade the performance of the proposed model by creating large variations in the observation counts across different cubes, leading to misestimated similarity scores.}

	\subsection{ Trade-off between computational cost and reconstruction accuracy}
	\label{app:N}
	{  The proposed method exhibits a trade-off between computational cost and reconstruction accuracy when scaling to larger datasets or more complex fields. FieldFormer is particularly well-suited for handling spatial-temporal continuous fields, which naturally result in a sparse attention map. This allows the model to maintain a relatively low model complexity and computational cost,  while achieving state-of-the-art reconstruction accuracy. However, when applied to more complex fields, the model adapts accordingly and the  percentage of the non-zero elements in the Tucker core tensor could increase, activating more learnable parameters. As a result, computational costs rise to ensure accurate reconstruction. Similarly, when scaling to larger datasets, more cubes are extracted to compute similarity scores, expanding the sparse attention map. This requires additional parameters to characterize the entire model, inevitably increasing the computational cost to maintain reconstruction accuracy. Nevertheless, the concern of computational overhead can be mitigated by implementing linear attention mechanisms \cite{flash}, which provide a more efficient alternative while preserving accuracy.}
	
\end{appendix}

\newpage 
\begin{IEEEbiography}[{\includegraphics[width=1in,height=1.25in,clip,keepaspectratio]{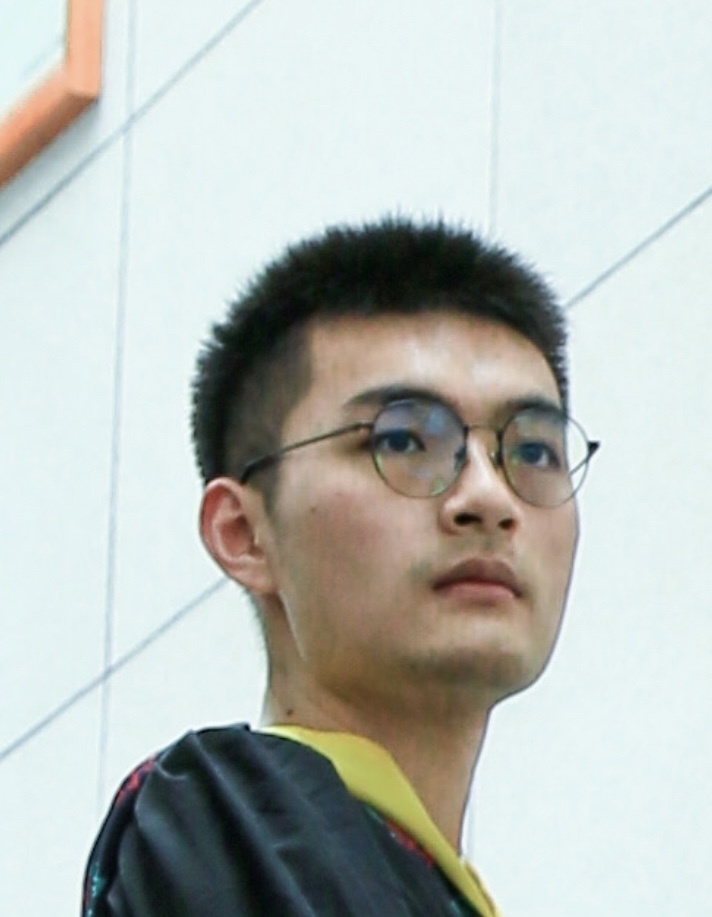}}]{Panqi Chen}
	 received the BSc degree from Xidian University, Xi'an, China, in 2022. He is currently working toward the PhD degree with the College of Information Science and  Electronic engineering, Zhejiang University, Hangzhou, China. His research interests include signal processing and machine learning.\end{IEEEbiography}
\begin{IEEEbiography}[{\includegraphics[width=1in,height=1.25in,clip,keepaspectratio]{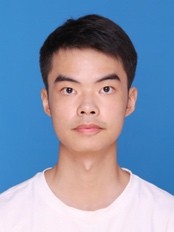}}]{Siyuan Li}
	 received the BSc degree in Electronic Science and Technology from Zhejiang University, Hangzhou, China. He is currently working toward the PhD degree with the College of Information Science and Electronic Engineering, Zhejiang University. His research interests include signal processing and machine learning.\end{IEEEbiography}

\begin{IEEEbiography}[{\includegraphics[width=1in,height=1.25in,clip,keepaspectratio]{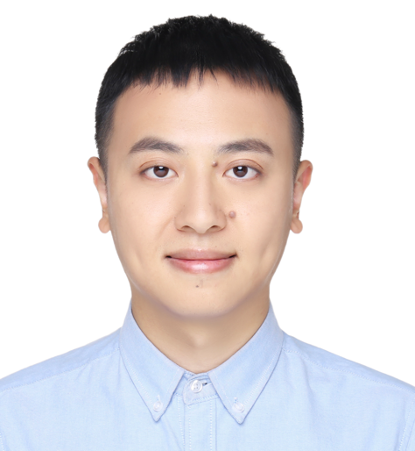}}]{Lei Cheng}(Member, IEEE) is currently ZJU Young Professor with the College of Information Science and Electronic Engineering, Zhejiang University, Hangzhou, China. He received the B.Eng. degree from Zhejiang University in 2013, and the Ph.D. degree from The University of Hong Kong in 2018. He was a Research Scientist in Shenzhen Research Institute of Big Data, The Chinese University of Hong Kong, Shenzhen, from 2018 to 2021. He is the author of the book “Bayesian Tensor Decomposition for Signal Processing and Machine Learning”, Springer, 2023. He was a Tutorial Speaker in IEEE ICASSP 2023, Invited Speaker in ASA 2024 and IEEE COA 2014. He is now Associate Editor for Elsevier Signal Processing and Young Editor for Acta Acustica. His research interests are in Bayesian machine learning for tensor data analytics and interpretable machine learning for information systems..\end{IEEEbiography}

\begin{IEEEbiography}[{\includegraphics[width=1in,height=1.25in,clip,keepaspectratio]{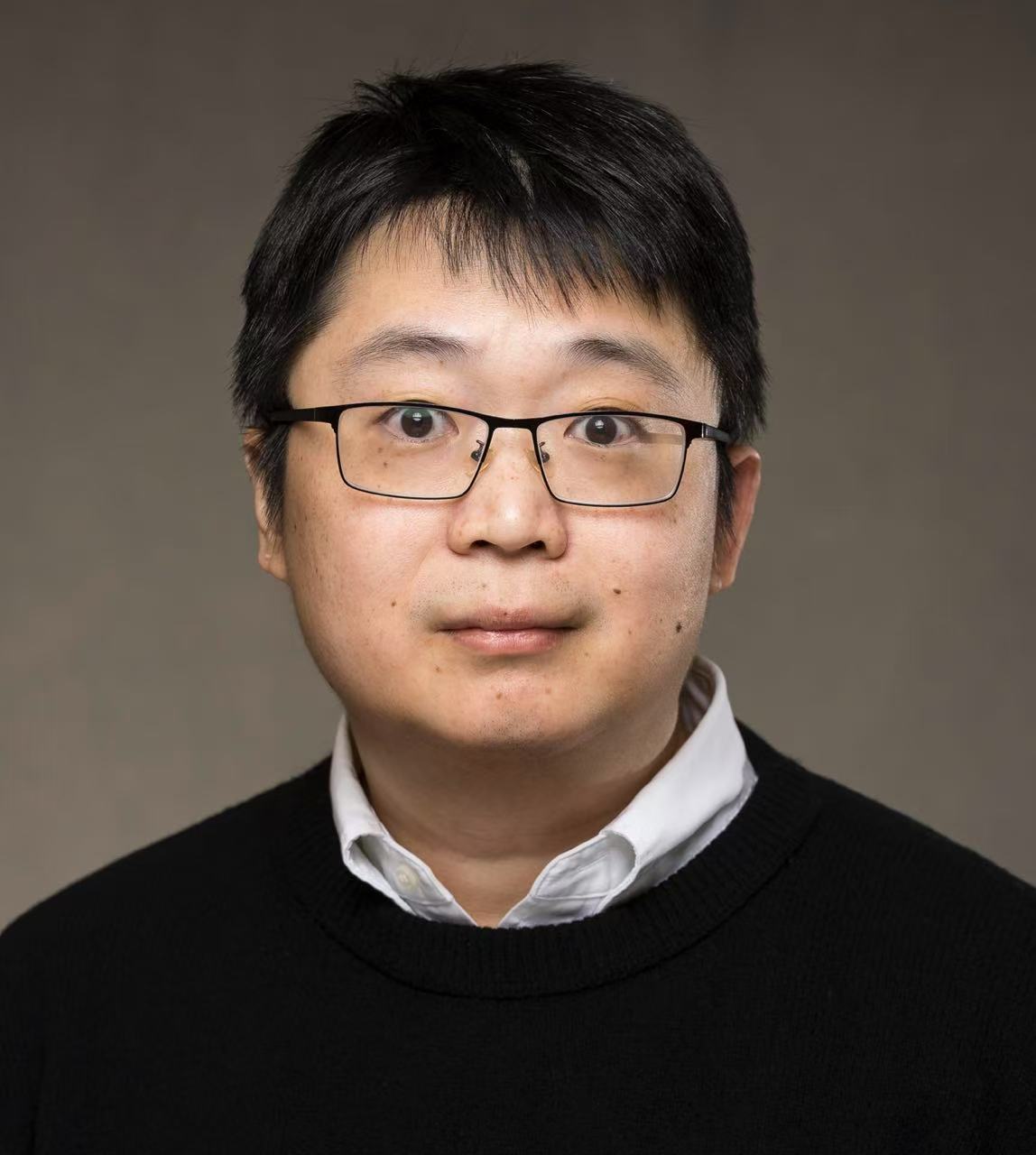}}]{Xiao Fu}
(Senior Member, IEEE) received the the Ph.D. degree in Electronic Engineering from The Chinese
University of Hong Kong (CUHK), Shatin, N.T., Hong Kong, in 2014. He was a Postdoctoral Associate with the
Department of Electrical and Computer Engineering, University of Minnesota, Minneapolis, MN, USA, from 2014
to 2017. Since 2017, he has been with the School of Electrical Engineering and Computer Science, Oregon State
University, Corvallis, OR, USA, where he is now an Associate Professor. His research interests include the broad
area of signal processing and machine learning. Dr. Fu received the 2022 IEEE Signal Processing Society (SPS) Best Paper Award and the 2022 IEEE SPS Donald
G. Fink Overview Paper Award. He is a recipient of the National Science Foundation (NSF) CAREER Award
in 2022, the College of Engineering Engelbrecht Early Career Award in 2023, and the University Promising Scholar Award in 2024.
He serves as a member of the IEEE SPS Sensor Array and Multichannel Technical Committee (SAM-TC) and the Signal Processing Theory and
Methods Technical Committee (SPTM-TC). He is currently a Suject Editor of SIGNAL PROCESSING and an Associate
Editor of IEEE TRANSACTIONS ON SIGNAL PROCESSING.\end{IEEEbiography}

\begin{IEEEbiography}[{\includegraphics[width=1in,height=1.25in,clip,keepaspectratio]{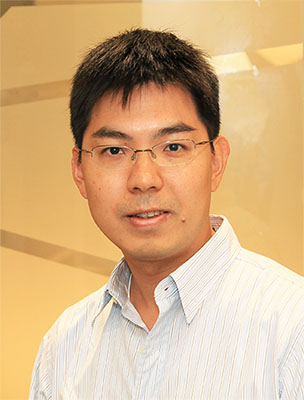}}]{Yik-Chung Wu}
 (Senior Member, IEEE)
received the B.Eng. (EEE) and M.Phil. degrees from The University of Hong Kong (HKU) in 1998 and 2001, respectively, and the Ph.D. degree from Texas A\&M University, College Station, in 2005. From 2005 to 2006, he was with Thomson Corporate Research, Princeton, NJ, USA, as a Member of Technical Staff. Since 2006, he has been with HKU, where he is currently as an Associate Professor. He was a Visiting Scholar at Princeton University in Summers of 2015 and 2017. His research interests include signal processing, machine learning, and communication systems. He served as an Editor for IEEE COMMUNICATIONS LETTERS and IEEE TRANSACTIONS ON COMMUNICATIONS. He is currently an Editor for IEEE TRANSACTIONS ON SIGNAL PROCESSING and Journal of Communications and Networks.\end{IEEEbiography}

\begin{IEEEbiography}[{\includegraphics[width=1in,height=1.25in,clip,keepaspectratio]{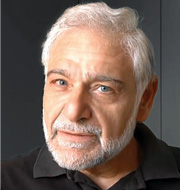}}]{Sergios Theodoridis}(Life Fellow, IEEE)
	is Professor Emeritus with the Department of Informatics and Telecommunications of the National and Kapodistrian University of Athens, Greece, and Director of Education of the HERON European center of excellence for AI and Robotics, Greece. 
	
	He is the author of the book “Machine Learning: From the Classics to Deep Networks, Transformers and Diffusion Models”, Academic Press, 3rd Ed., 2025, the co-author of the best-selling book “Pattern Recognition”, Academic Press, 4th Ed. 2009, the co-author of the book “Introduction to Pattern Recognition: A MATLAB Approach”, Academic Press, 2010.  
	
	He is the co-author of seven papers that have received Best Paper Awards including the 2014 IEEE Signal Processing Magazine Best Paper Award and the 2009 IEEE Computational Intelligence Society Transactions on Neural Networks Outstanding Paper Award. 
	
	He has received an honorary doctorate degree (D.Sc) from the University of Edinburgh, UK, in 2023. He is the recipient of the 2021 IEEE Signal Processing Society (SPS) Norbert Wiener Award, the 2017 EURASIP Athanasios Papoulis Award, the 2014 IEEE SPS Carl Friedrich Gauss Education Award and the 2014 EURASIP Meritorious Service Award. He has served as a Distinguished Lecturer for the IEEE SP as well as the Circuits and Systems societies. 
	
	He has served as Vice President IEEE Signal Processing Society, as President of the European Association for Signal Processing (EURASIP), as a member of the Board of Governors for the IEEE Circuits and Systems (CAS) Society,  and as the chair of the IEEE SPS awards board.
	
	He is Fellow of IET, a Corresponding Fellow of the Royal Society of Edinburgh (RSE), a Fellow of EURASIP and a Life Fellow of IEEE.
	\end{IEEEbiography}

\end{document}